\documentclass[11pt,letterpaper]{article}
\pdfoutput=1
\usepackage{jheppub}
\usepackage{scalefnt}
\usepackage{xfrac}
\usepackage{caption}
\usepackage{subcaption}

\PassOptionsToPackage{jheppub}{hyperref}

\makeatletter
\everymath{\if@display\else\thickmuskip=4mu plus 4mu\fi}
\makeatother

\newcommand{\beq}{\begin{equation}}
\newcommand{\eeq}{\end{equation}}
\newcommand{\bea}{\begin{eqnarray}}
\newcommand{\eea}{\end{eqnarray}}

\def\ocal{{\cal O}}
\def\rcal{{\cal R}}

\def\pa{\partial}
\newcommand{\vev}[1]{\left\langle#1\right\rangle}
\newcommand{\eqn}[1]{Eq.~{\hspace{-2pt}}(\ref{#1})}
\newcommand{\eqns}[2]{Eqs.~{\hspace{-2pt}}(\ref{#1}),\,(\ref{#2})}
\newcommand{\figr}[1]{Fig.~\hspace{-2pt}\ref{#1}}
\newcommand{\secn}[1]{Sec.~\hspace{-2pt}\ref{#1}}
\newcommand{\tabby}[1]{Table~\hspace{-2pt}\ref{#1}}
\newcommand{\tabbys}[2]{Tables~\hspace{-2pt}\ref{#1}\,\&\,\ref{#2}}

\newcommand{\reference}[1]{Ref.~\hspace{-2pt}\cite{#1}}

\newcommand{\half}{\frac{1}{2}}

\newcommand{\abar}{{\overline a}}
\newcommand{\bbar}{{\overline b}}
\newcommand{\betabar}{{\overline{\beta}}}
\newcommand{\alphabar}{{\overline \alpha}}

\newcommand{\uv}{ {}^{(\mathrm{uv})}\!}
\newcommand{\wt}[1]{\widetilde{#1}}
\def\Tr{{\rm Tr }}

\def\mn{{\mu\nu}}

\def\xip{{\xi'}}
\def\delvar{{\delta\varphi}}

\author[a,1]{Martin B Einhorn}\note{Also, \it{Michigan Center for 
Theoretical Physics, Ann Arbor, MI 48109.}} 
\author[a,b]{and D R Timothy Jones}
\affiliation[a]{Kavli Institute for Theoretical Physics,\\ University of California,
Santa Barbara, CA 93106-4030, USA}
\affiliation[b]{Dept. of Mathematical Sciences,\\ University of Liverpool, 
Liverpool L69 3BX, UK}

\emailAdd{meinhorn@umich.edu}
\emailAdd{drtj@liv.ac.uk}

\title{Induced Gravity II: Grand Unification}

\abstract{As an illustration of a renormalizable, asymptotically-free
model of induced gravity, we consider an $SO(10)$ gauge theory
interacting with a real scalar multiplet in the adjoint
representation. We show that dimensional transmutation can occur,
spontaneously breaking $SO(10)$ to $SU(5){\otimes}U(1),$ while
inducing the Planck mass and a positive cosmological constant, all
proportional to the same scale $v$. All mass ratios are functions of
the values of coupling constants at that scale. Below this scale (at
which the Big Bang may occur), the model takes the usual form of
Einstein-Hilbert gravity in de~Sitter space plus calculable
corrections. We show that there exist regions of parameter space in
which the breaking results in a local minimum of the effective action 
giving a positive dilaton $(\hbox{mass})^2$ from two-loop
corrections associated with the conformal anomaly. Furthermore, unlike
the singlet case we considered previously, some minima lie within the
basin of attraction of the ultraviolet fixed point. Moreover, the
asymptotic behavior of the coupling constants also lie within the
range of convergence of the Euclidean path integral, so there is hope
that there will be candidates for sensible vacua. Although open
questions remain concerning unitarity of all such renormalizable
models of gravity, it is not obvious that, in curved backgrounds such
as those considered here, unitarity is violated. In any case, any
violation that may remain will be suppressed by inverse powers of the
reduced Planck mass.}

\keywords{Renormalization Group, Models of Quantum Gravity, GUT}
\arxivnumber{1602.06290}

\begin{document}
\maketitle
\vspace{-10mm}
\listoffigures
\medskip
\hrule
\vfill
\pagebreak

\section{Introduction}
\label{sec:intro}

The standard model (SM) cannot be perturbatively ultra-violet complete
simply because of the presence of a U(1) gauge coupling, inevitably
leading to a Landau pole. However the SM, when made supersymmetric, or
by inclusion of other suitably chosen light states, does suggest the
possibility of a gauge unification scale $M_X$ of around $10^{16}$~GeV,
corresponding to new physics based on a gauge group containing as a
subgroup $SU(3){\otimes}SU(2){\otimes}U(1)$. Models based on this idea
typically involve proton decay mediated by particles with unification
scale masses; predicting rates close to if not violating experimental
limits. The relationship between $M_X$ and the Planck scale $M_P\sim
10^{19}$~GeV (or the reduced Planck mass or string scale
$M_P/\sqrt{8\pi},)$ has long been a source of inquiry in the context of
efforts to construct an ultimate theory.

One point of view is that the ratio $M_X/M_P$ being $O(10^{-3})$ is a 
good thing in rendering perturbation theory valid at $M_X$; another is 
that the existence of the two nearby scales is un-aesthetic, and the low
energy theory should be modified so as to move $M_X$ up to $M_P$. In 
either case, the question of the nature of the ultimate theory remains.
One approach to this question is string theory. We follow an older path,
that of renormalizable quantum field theory (QFT), including
gravity~\cite{Stelle:1976gc}. This theory of gravity, sometimes called
$``R^2 "$ gravity or ``higher-derivative" gravity, has another
attractive feature inasmuch as it is asymptotically free 
(AF)~\cite{Fradkin:1981hx, Fradkin:1981iu}. Under certain 
circumstances, these properties may be extended to include matter in 
its usual form of scalar, vector, and fermion fields, corresponding to 
spins $(0, 1, 1/2).$ 

Since this paper is a sequel to others~\cite{Einhorn:2015lzy, 
Einhorn:2014gfa}\ along these same lines, we limit describing the
motivation for this work to a few other introductory remarks. In
addition to renormalizability and AF for all couplings, we, as do the
authors of \reference{Salvio:2014soa}, restrict our attention to such
extensions that are classically scale invariant. This bequeathes certain
 naturalness properties to the theory that are essential to avoid issues
of fine-tuning, even in the presence of the breaking of scale invariance
by the conformal anomaly~\cite{Bardeen:1995kv}. It is also aesthetically
attractive in that there are no elementary masses to be accounted for,
and all mass scales must ultimately be due to dimensional transmutation
(DT), whether perturbatively~\cite{Coleman:1973jx}, or
nonperturbatively, as in Yang-Mills theory~\cite{Yang:1954ek} or
massless QCD\footnote{For some recent speculations about strong coupling
in this context, see \reference{Holdom:2015kbf}. As we have remarked
previously~\cite{Einhorn:2014gfa}, even classically, a
$1/q^4$-propagator corresponds to a linearly growing potential, which
would therefore be confining.}. We shall focus exclusively on the
perturbative scenario. 

In previous work~\cite{Einhorn:2015lzy, Einhorn:2014gfa}, we 
considered the simplest possible extension of renormalizable gravity, 
viz., to the inclusion of a single, real scalar field. We showed 
that such a model can simultaneously generate by DT a scalar vacuum 
expectation value (VEV) and nonzero scalar curvature $R$. Moreover 
the theory has a region of parameter space containing an ultra-violet 
stable fixed point (UVFP) for coupling constant ratios and is AF in 
all its coupling constants. Unfortunately, however, the region of 
parameter space corresponding to DT and a ``right-sign'' Einstein term 
$(\xi{>}0)$ was disjoint from the basin of attraction of the UVFP: 
starting from the DT region, the couplings did not flow to the UVFP.

In this paper, we extend the results of \reference{Einhorn:2015lzy}\ 
to the case in which the matter sector includes non-Abelian gauge 
interactions and non-singlet scalars and fermions for which all the 
couplings are AF. We show that, not only does the same DT phenomenon 
occur, but the disappointing outcome mentioned above does not hold; 
this time there {\it is\/} a region of parameter space such that both 
DT occurs at a local minimum from which the couplings flow to the 
UVFP. Moreover, both $M_P$ {\bf and\/} $M_X$ can be understood in 
terms of the scalar VEV. 

In flat space, if Yukawa couplings are AF, then they usually fall faster
than the quartic scalar couplings. There is no guarantee, however, that
they are negligible at the DT scale. Our goal in the present effort is
not to obtain a completely realistic model but to determine whether we
can find any model of this type that realizes all our many 
constraints\footnote{Models of GUTs within renormalizable  gravity were
considered long ago~\cite{Buchbinder:1989ma,  Buchbinder:1989jd}, but
that work did not consider induced gravity  or any of the constraints
that we impose other than AF in all couplings. Induced gravity in models
of GUTs have been previously considered, e.g., in
\reference{CervantesCota:1994zf}, but not in the context of
renormalizable  gravity with dimensional transmutation.},  so, for
present purposes, we shall ignore possible Yukawa couplings. 

To summarize our goals: we seek a model that 

(1)~is AF for values of the couplings that insure
convergence of the EPI at sufficiently high scales, 

(2)~undergoes DT at some scale, with a locally stable minimum, 

(3) is such that a portion of the range of couplings satisfying the 
preceding constraint lies within the basin of attraction of the UVFP, 
so that the couplings run from DT solutions to the UVFP. This is 
where our previous attempts failed.
 
We shall, in fact, be successful in all these goals.

\section{Classically Scale Invariant Gravity}\label{scaleinv}

The basic framework for this paper is classically scale invariant 
quantum gravity, defined by the Lagrangian  
\beq\label{eq:hoaction} 
S_{ho}=\int d^4x\sqrt{g}\left[ \frac{C^2}{2a}+\frac{R^2}{3b}+cG \right],
\eeq 
where $C$ is the Weyl tensor and $G$ is the Gauss-Bonnet 
term\footnote{We work in Euclidean spacetime throughout with the 
curvature conventions given in \reference{Einhorn:2014gfa}}. Just about 
the simplest imaginable scale invariant theory involving gravity and 
matter fields consists of the above, coupled to a single scalar field 
with a $\lambda\phi^4$ interaction and non-minimal gravitational 
coupling $\xi R \phi^2$. In recent papers~\cite{Einhorn:2014gfa, 
Einhorn:2015lzy}, we argued that even this matter-free theory can
undergo  dimensional transmutation (DT) \` a la 
Coleman-Weinberg~\cite{Coleman:1973jx}, leading to effective action 
extrema\footnote{We use the term ``extrema" to refer to stationary 
points generally, not just maxima and minima.} with nonzero values for 
$\vev{R}$ and $\vev{\Phi}$ (or $\vev{T_2}).$ However, the extrema are
unstable and consequently unacceptable. It is important to emphasize 
that, as with the original treatment~\cite{Coleman:1973jx} of scalar 
electrodynamics, we restrict ourselves to DT that can be demonstrated 
perturbatively; in other words, for values of the relevant dimensionless
couplings such that neglect of non-leading quantum corrections can be 
justified. In this paper we shall take the matter action to be that of a
gauge field of a simple group, with a real scalar field in the adjoint 
representation\footnote{The generalization to a semi-simple gauge group 
is straightforward, but $U(1)$ factors are not permitted, since an 
abelian gauge coupling cannot be asymptotically free.}: 
\beq\label{eq:mataction}
S_m=\int d^4x \sqrt{g}\left[\frac{1}{4}\Tr[F_{\mu\nu}^2]+ 
\half\Tr[(D_\mu\Phi)^2]-\frac{\xi \Tr [\Phi^2]}{2} R +V_J(\Phi)\right],
\eeq
where $\Phi = \sqrt{2} T^a \phi^a$ with $\phi_a$ real, 
$D_\mu\Phi \equiv \pa_\mu\Phi +i g[A_\mu,\Phi],$ 
$A_\mu \equiv \sqrt{2} T^a A_\mu^a,$ and 
$F_{\mu\nu} \equiv \pa_\mu A_\nu-\pa_\nu A_\mu + i g [A_\mu,A_\nu].$ 
By definition, the 
generators $T^a$ are Hermitian and conventionally taken to be in 
the defining or fundamental representation of the group, normalized so 
that $\Tr[T^aT^b] = \delta^{ab}/2.$ Thus, with our conventions, 
$\Tr[\Phi^2] = \sum(\phi_a)^2.$ We take the potential to be 
\begin{subequations}
\label{eq:potential}
\begin{align}
\label{eq:potv}
V_J (\Phi) &\equiv \frac{h_1}{24} T_2^{\,2}+\frac{h_2}{96} T_4,
\ \mathrm{or}\\
\label{eq:potvv}
V_J (\Phi) &\equiv \frac{h_3}{24} T_2^{\,2}+\frac{h_2}{96} \wt{T}_4,\ 
\mathrm{where}\ \wt{T}_4\equiv\left[T_4-\frac{1}{d_T}T_2^{\,2}\right],
\end{align}
\end{subequations}
where $T_n {\equiv} \Tr[\Phi^n]$ and $d_T$ is the dimension of the 
fundamental representation $T^a\!.$
For $SO(N)$ (and $SU(N)$),
$d_T{=}N$ for their fundamental representations.
The relation between the couplings in the two 
expressions is $h_3 {\equiv} h_1 {+} h_2/(4d_T).$ 
It can be easily shown that 
$T_4 {\geq} T_2^{ 2}/d_T,$ so that $\wt{T}_4 \geq 0.$ 

Classically, for the potential to be bounded below, one must have 
$h_2{>}0$ and $h_3{>}0.$ In the QFT, it is unclear at what scale this is 
required of the renormalized couplings $\{h_2(\mu), h_3(\mu)\}$ or, 
equivalently, that this classical requirement is necessary for the effective 
action to be bounded below. In fact, because of AF, the classical form of 
the renormalized action is an increasingly good approximation the larger 
the scale $\mu$ so these constraints are reliable for $\mu$ sufficiently 
large\footnote{Precisely the same constraint results from 
demanding convergence of the path integral. See 
\secn{sec:constraints}.}. As the scale $\mu$ decreases, one must 
determine from the renormalization-group-improved effective action how 
far down in the infrared (IR) direction these inequalities will continue to 
remain necessary, assuming that it remains within the realm of a 
perturbative calculation.
 
To \eqn{eq:mataction}, we shall add a certain number of 
massless fermions in representations yet to be specified. 
For simplicity, we shall ignore possible Yukawa interactions. Without 
gravitational interactions, it was remarked long ago~\cite{Cheng:1973nv} 
that, so long as they are themselves asymptotically free, 
Yukawa couplings vanish more rapidly in the UV than gauge couplings 
and scalar self-couplings, so their presence does not affect the 
asymptotic behavior of the other couplings. This conclusion survives 
the inclusion of the gravitational couplings in the cases we shall 
consider, although the sign of their contribution does in fact act so as to 
make the Yukawa couplings vanish {\bf less} rapidly\footnote{With the 
original form of the beta-functions given, e.g., in 
\reference{Buchbinder:1992rb}, they vanish {\bf more} rapidly. As we described 
in \reference{Einhorn:2015lzy}, we have adopted the alternative 
beta-functions given in~\reference{Salvio:2014soa}.}. They could in principle 
affect the equations for DT in important ways, but to keep things simple, 
we shall assume they can be neglected down to the DT scale.

\section{Beta-functions for an SO(N) model and asymptotic freedom}
\label{sec:betas}

One attractive property of renormalizable gravity defined by \eqn{eq:hoaction} 
is that it is asymptotically free (AF), and this property can be extended to 
include a matter sector with an asymptotically free gauge theory, or even 
a non-gauge theory, such as the ones considered 
previously~\cite{Einhorn:2014gfa,Einhorn:2015lzy}. 
This can be seen as follows: At one-loop order, the gauge coupling
$g$ and the gravitational couplings $a$ and $c$ do not mix with 
other couplings. In the general case, their $\beta$-functions 
are\footnote{We suppress throughout a factor 
$1/(16\pi^2)$ from all one-loop $\beta$-functions.} 
\begin{align}
\label{eq:betagac}
\beta_{g^2}&=-b_g (g^2)^2, &\qquad \beta_a&=-b_2 a^2,&\qquad \beta_c&=-b_1,\\
\label{eq:bgb1b2}
b_g&=2(\frac{11}{3}C_G -\frac{2}{3}T_F -\frac{1}{6}T_S), &\qquad 
b_2&= \frac{133}{10}+ N_a, &\qquad 
b_1&=\frac{196}{45}+ N_c,
\end{align} 
where $N_a = \left[N_0+3N_F+12N_V\right]\!/60$ and 
$N_c = \left[N_0+\frac{11}{2}N_F+62N_V\right]\!/360$. 
Here, $N_0$ represents the number of real scalars; 
$N_V,$ the number of massless vector bosons;
$N_F,$ the number of Majorana or Weyl fermions. Our Lie algebra 
conventions are summarized in Appendix~\ref{sec:liealgebra}. 
Since $b_2{>}0,$ the coupling $a$ is always AF; we may estimate its 
rate of decline by noting the form of $N_a$ above. Typically, $N_{a,c}$ 
are dominated by vector bosons and fermions, since scalars are down 
from vectors by a factor of 12. In the $SO(N)$ model that we 
consider below, with a single, real, adjoint scalar, 
$N_a = 13N(N{-}1)/120+N_F/20\/$. 
As we shall explain shortly, it turns out that there are AF solutions 
for the scalar couplings only for $N \geq 9,$ so 
$N_a \geq 39/5+N_F/20\/$ and 
$b_2 \geq 211/10+N_F/20$. (Obviously, this lower bound grows 
quadratically with increasing $N$.)

The evolution of the coupling $b$ is more complicated:
\beq
\label{eq:betab}
\beta_b\equiv -a^2 b_3(x,\xip),\quad 
b_3(x,\xip)\equiv \left[\frac{10}{3}-5 x+ \left(\frac{5}{12} 
+\frac{3\xip^2 N_0}{2}\right)x^2\right],\\
\eeq
where $x \equiv b/a,$ and we have introduced $\xip \equiv \xi+1/6.$ (Whereas
$\xi = 0$ for minimal coupling, $\xip = 0$ for conformal coupling.)
Thus, $b$ mixes with the couplings $a$ and $\xip,$ and 
$\beta_\xip$ depends on the matter self-couplings. Therefore, unlike 
$a,$ the evolution of $b$ is sensitive to other features of the model. 

For reasons explained in \reference{Einhorn:2015lzy}, we adopt the 
beta-functions of Salvio and Strumia~\cite{Salvio:2014soa}, which differ 
for matter couplings from earlier results~\cite{Buchbinder:1992rb}. For 
the $SO(N)$ case with a single adjoint scalar field, the remaining 
beta-functions are\footnote{The flat space beta-functions for 
$\{ \beta_{h_1}, \beta_{h_2} \}$ can be found in \reference{Cheng:1973nv}.}
\begin{subequations}\label{eq:betas}
\begin{align}
\begin{split}
\beta_{h_1}& = \frac{1}{3}\left(\frac{N (N{-}1)}{2}+8\right) h_1^2 +
\frac{2 N{-}1}{12} h_1 h_2+\frac{1}{32}h_2^2-6(N{-}2)h_1g^2 +\\
&\hskip20mm 27g^4 +3\Delta\beta_1 + h_1 \Delta\beta_2,\label{eq:betah1}
\end{split}\\
\beta_{h_2}& = 4 h_1 h_2+ \frac{2 N{-}1}{24} h_2^2
-6 (N{-}2) h_2 g^2+36(N{-}8) g^4+h_2 \Delta\beta_2, 
\label{eq:betah2} \\
\label{eq:deltabetas}
\Delta\beta_1& = a^2\Big(\xip{-}\frac{1}{6} \Big)^{\!2} \left(5+ 9x^2 \xip^2 \right),\quad
\Delta\beta_2 = a \left(5-18x\xip^2\right),\\
\label{eq:betaxip}
\beta_\xip& = \xip\left(\Big(\frac{N(N{-}1)+4}{6}\Big)h_1
+\frac{2N{-}1}{24}h_2 -3 (N{-}2) g^2\right)
+\Delta\beta_\xip,\\
\Delta\beta_\xip& = a\Big(\xip{-}\frac{1}{6}\Big)\!\! \left(\frac{10}{3 x}-\frac{3}{2} \xip(2\xip{+}1) x\!\right)=\Big(\xip{-}\frac{1}{6}\Big)\!\! \left(\frac{10a^2}{3 b}-\frac{3}{2} \xip(2\xip{+}1) b\!\right).
\end{align}
\end{subequations}

It is interesting that the gravitational contribution to $\beta_\xip$, viz.\!
$\Delta\beta_\xip$, vanishes for {\it minimal\/} coupling, whereas the 
matter contributions vanish for {\it conformal\/} coupling, 
 about which we shall have more to say shortly. We want to 
examine the possibility of obtaining a theory in which all of the couplings 
are AF. We must 
demand $b_g{>}0$, so that the gauge coupling is AF. In a certain sense, the evolution 
of the two couplings $a$ and $g^2$ control the behavior of the other couplings. To 
see this, it is useful to rescale the other couplings by one of these two and 
to express their beta-functions in terms of these ratios; since neither coupling 
vanishes at any finite scale, we may choose to rescale by either one. 
In theories without AF gauge couplings, one must choose $a,$ as we did 
in our previous papers. In gauge models, it is more 
convenient~\cite{Buchbinder:1992rb} to rescale by $\alpha \equiv g^2$ 
instead, replacing the conventional running parameter $dt=d\ln\mu$ by 
$du=\alpha(t) dt.$ This enables us to easily investigate the 
impact of gravitational corrections on the flat-space beta-functions. 
Thus we introduce rescaled couplings:
\beq\label{eq:rescaled}
z_1 \equiv \textstyle{\Large{\sfrac{h_1}{\alpha}}},\quad 
z_2 \equiv \textstyle{\Large{\sfrac{h_2}{\alpha}}},\quad 
z_3 \equiv \textstyle{\Large{\sfrac{h_3}{\alpha}}},\quad 
\abar \equiv \textstyle{{\Large\sfrac{a}{\alpha}}},\quad 
\bbar\equiv \textstyle{{\Large{\sfrac{b}{\alpha}}}}.
\eeq 
As we shall see, because of the nature of the symmetry breaking of the 
$SO(N)$ group in this model, it is usually simpler 
to use the pair $\{z_2, z_3\}$ than $\{z_1, z_2\}$. Of course, 
$x\equiv b/a=\bbar/\abar,$ and need not be rescaled. Note that $\xip$ 
is not rescaled\footnote{\label{foot:xipuv}For asymptotic freedom, we 
only require $\xip\to\xip\uv,$ some finite constant. In that event, we 
could trivially replace $\xip$ by $\xi^{''}\!\equiv\xip-\xip\uv,$ which 
approached zero. Thus, so long as $\xip$ approaches \emph{any\/} 
finite constant asymptotically, the theory can be said to be AF. We 
shall also show however that $\xip\uv$ is naturally extremely small 
but nonzero, so that such theories are never asymptotically 
conformal.}. If $\xip$ and the ratios 
$\{\abar, \bbar, z_2, z_3\}$ approach a finite UVFP, then the original 
couplings $\{\alpha, a, b, h_1, h_3\}$ will all be AF. The rescaled 
beta-functions, $\betabar_{\lambda_i}$ correspond to $d\lambda_i/du.$ 
Noting that $\beta_{h_3} = \alpha^2(\betabar{}_{z_3}-b_g z_3),$ and 
$\beta_{h_2} = \alpha^2(\betabar{}_{z_2}-b_g z_2),$ we find 
\begin{subequations}
\label{eq:betabars}
\begin{align}
\label{eq:betaabar}
\betabar_{\abar} &= \abar \left(b_g-\abar b_2\right),\\
\label{eq:betabbar}
\betabar_\bbar-b_g\bbar&=-\abar^2 b_3(x,\xip)=
\left[{-}\frac{10}{3}\abar^2 + 5 \abar \bbar- 
\left(\frac{5}{12} +\frac{3 N(N{-}1)\xip^2}{4}\right)\bbar^2\right],\\
\label{eq:betaxbar}
\betabar_x-b_2 x \abar& = -b_3(x,\xip) \abar = 
\abar \left[-\frac{10}{3}+5 x-
\frac{x^2}{12}\left(5+9 N(N{-}1)\xip^2\right)\right],\\
\label{eq:betaz2bar}
\betabar{}_{z_2}-b_g z_2& = 36(N{-}8)+\frac{2N^2{-}N{-}24}{24N}z_2^2+ 4 z_3 z_2-
6(N{-}2)z_2+\overline{\Delta\beta}_2 z_2,\\
\begin{split}
&\hskip-17mm\betabar{}_{z_3}-b_g z_3 = 
\frac{36(N{-}2)}{N} +\frac{N(N{-}1)+ 16}{6} z_3^2
 + \frac{N^2{-}4}{48N^2} z_2^2+ \frac{N^2{-}4}{12N} z_3z_2-\hskip2mm\ \cr
& \hskip20mm 6(N{-}2)z_3+\overline{\Delta\beta}_2 z_3
+3\overline{\Delta\beta}_1,
\label{eq:betaz3bar}\\
\end{split}\\
\label{eq:deltabeta1bar}
\overline{\Delta\beta}_1& =
\abar^2 \Big(\xip{-} \frac{1}{6}\Big)^{\!2}\left(5+9x^2 \xip^2\right)=
\Big(\xip{-} \frac{1}{6}\Big)^{\!2} 
\left(5 \abar^2+9 \bbar^2 \xip^2\right),\\
\label{eq:deltabeta2bar}
\overline{\Delta\beta}_2& =\abar \left(5-18 x \xip^2\right)=
5 \abar-18 \bbar \xip^2,\\
\label{eq:betaxipbar}
\betabar_\xip& = \xip \left[\frac{N^2{-}4}{24N} z_2+\frac{N(N{-}1)+ 4}{6}z_3-3(N{-}2)\right]
+\overline{\Delta\beta}_\xip,\\
\label{eq:deltabetaxipbar}
\overline{\Delta\beta}_\xip& =\abar \left(\!\xip{-}\frac{1}{6}\right) 
\left[\frac{10}{3 x}-\frac{3}{2} \xip(2\xip{+}1) x\right]\!=\!
\left(\!\xip{-}\frac{1}{6}\right)\!\! 
\left[\frac{10 \abar^2}{3 \bbar}-\frac{3}{2} \xip(2\xip{+}1) \bbar\right]\!.\!
\end{align}
\end{subequations}
All dependence on $\alpha$ has disappeared. For historical 
reasons, we retained the ratio $x\equiv b/a=\bbar/\abar$, but it turns out 
that, to search for candidates for UVFPs, it is usually better to work with 
$\bbar.$ Although redundant, we have given both $\betabar_\bbar$ and 
$\betabar_x$ and expressed the gravitational corrections 
$\overline{\Delta\beta}_k, (k=1,2,\xip)$ in two alternative forms, each of 
which is useful in different contexts. We shall see shortly that 
$\bbar\to \bbar\uv\sim O\left(b_g\right)$,so that 
$\abar\uv/ \bbar\uv \sim \ocal(1/ b_2) \ll 1.$ Inversely, 
$x=\bbar/\abar \to x\uv=\bbar\uv/\abar\uv \sim \ocal(b_2) \gg 1.$ 

Notice that the gravitational corrections $\{ \overline{\Delta\beta}_1,
\overline{\Delta\beta}_2 \}$ do not depend upon $N$, so that the
dependence of $\betabar{}_{z_3}\/$ and $\betabar{}_{z_2}\/$
on $N\/$ is determined by the non-gravitational sector. We have
shown that, without the gravitational couplings, $SO(N)$ can in principle
have asymptotically free scalar couplings only for\footnote{In
\reference{Cheng:1973nv}, it was stated that $N = 8$ is also possible,
but that resulted from the approximation $b_g = 0$, where $b_g$ is the
one-loop gauge beta-function coefficient (explicitly given in the next
section.) In fact, in this class of models, asymptotic freedom 
mandates that $b_g \geq 1/6.$} $N \geq 9$. (Similarly, $SU(N)$ with an
adjoint scalar is required to have $N\ge 7$.) These conclusions remain
unaffected by including the gravitational interactions. As mentioned
earlier, Yukawa couplings may usually be added without affecting the
asymptotic behavior of the gauge or gravitational couplings 
$\{\alpha, a, b, \xip\}$, so long as they themselves are AF.

The challenge now is to determine whether or not these 
beta-functions in \eqn{eq:betabars} have at least one finite, UV-stable, 
fixed point (FP) in all the parameters. In fact, substantial progress can be
made in this simple model for arbitrary values of $N \geq 9.$ In the remainder 
of this section, we discuss the general properties of a potential 
UVFP. In fact, we will show that the UVFP in $\{\abar,\bbar,\xip\}$ can, to a 
good approximation, be determined analytically, and further that, to 
determine the UVFP in $\{z_2, z_3\},$ we need only find the UVFP for their 
flat-space beta-functions with a gravitationally modified factor for 
$b_g.$\footnote{Readers interested only in seeing the results for 
$SO(10)$ may safely skip forward to the next section.}

\begin{figure}[htbp]
\centering
\setlength\fboxsep{4pt}
\fbox{\includegraphics[width=4.75in]{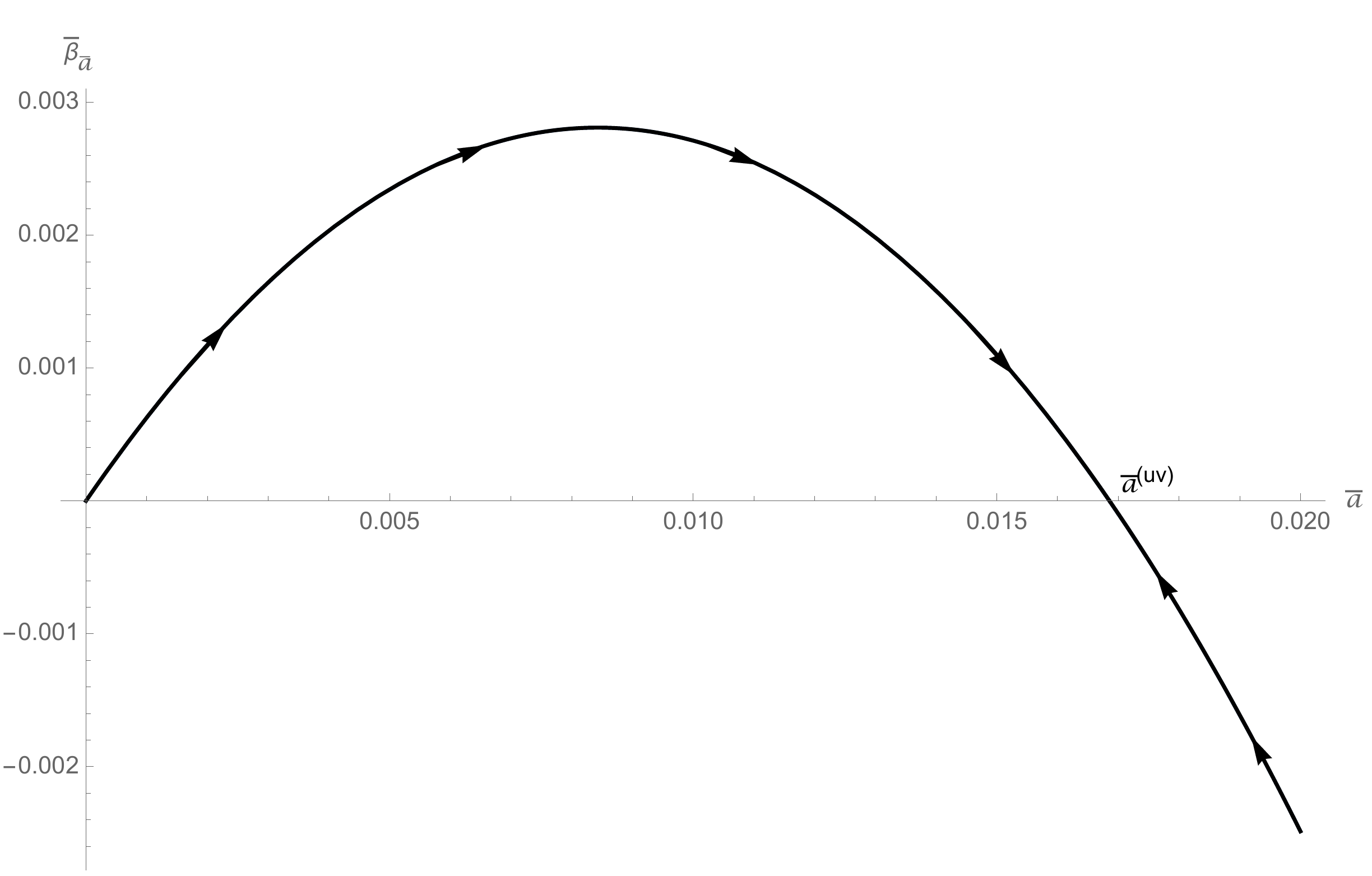}}
\caption{$\beta_\abar$ showing its UVFP at $\abar\uv$. } 
\label{fig:betaabar}
\end{figure}
The UV behavior of $\abar,$ \eqn{eq:betaabar}, is easily discerned since, like 
$\alpha$ and $a$, it does not mix with other couplings at one-loop order. In 
\figr{fig:betaabar}, we plot this beta-function\footnote{The actual numbers in 
\figr{fig:betaabar} correspond to an example that will be used in subsequent 
figures and tables. An illustration of running $\abar(u)$ from the DT-scale 
toward its UVFP is given in \figr{fig:runningabar}.}, showing its UVFP at 
$\abar\uv=b_g/b_2{>}0.$ (Referring to \eqn{eq:bgb1b2}, we see that $b_2$ 
is always positive; we must require $b_g{>}0$ for the gauge coupling to be 
AF.) If this were the only coupling in the model, $\abar\uv$ would be the 
dividing line between two phases. That is not the case here. Assuming 
that we find a UVFP, $\abar\uv$ is simply one of its coordinates in the 
five-dimensional space of ratios $\{\abar,\bbar,\xip,z_2,z_3\}$. 

Nevertheless, because its beta-function is independent of the other couplings, 
the running of $\abar$ can be understood easily. As the coupling $\abar(u)$ 
runs from near the UVFP toward lower energy scales, $\abar(u)$ increases if 
it starts from $\abar{>}\abar\uv.$ On the other hand, if it starts at a value 
$\abar{<}\abar\uv,$ then it decreases as the scale decreases. In the first case, 
$a(u){>}\abar\uv \alpha(u)$, so gravitational interactions are 
becoming relatively \emph{stronger} than gauge interactions; in the second 
case, $a(u){<}\abar\uv \alpha(u)$, so gravitational interactions are becoming 
relatively \emph{weaker} than gauge interactions. In both cases, $a(u)$ and 
$\alpha(u)$ are increasing, but there will be no breakdown of perturbation 
theory unless either gravitational interactions or gauge interactions actually 
become strong. The alternative, the one explored in this paper, is that DT 
occurs before strong interactions set in.

A~priori, $b_g{>}0$ could take values over a large range. For reasons to be 
explained in greater detail in \secn{sec:catch}, it seems that $b_g\sim\ocal(1).$
The reason is the requirement that the scalar couplings be AF, to be discussed 
further in \secn{sec:catch}. As a result, $\abar\uv\lesssim \ocal(10^{-2}).$ E.g.,
in the $SO(10)${-}case discussed beginning in \secn{sec:so10}, we find 
$\abar\uv$ in the narrow range $0.015\lesssim\abar\uv\lesssim0.019.$


Another implication is that $\betabar_\xip \to \ocal(b_g/b_2^2) \lll 1,$ so that 
$\xip\uv$ will be nearly conformal but never exactly zero\footnote{Of course, 
conformal or Weyl gravity is assigned different counterterms in an 
attempt to enforce conformal symmetry. It is unclear whether this is
truly consistent.}. This can be seen as follows: As remarked in 
footnote~\hbox{\ref{foot:xipuv}}, $\betabar_\xip,$ \eqns{eq:betaxipbar}{eq:deltabetaxipbar}, vanishes for neither conformal nor minimal 
coupling. For conformal coupling $\xip = 0$ $(\xi = {-}1/6)$, the contribution 
in~\eqn{eq:betaxipbar} that is independent of gravitational corrections vanishes. This is 
the familiar property that, in a QFT in a fixed, background gravitational field, a free 
massless scalar field having $\xip = 0$ is classically conformally invariant. This has 
been conjectured to remain true if scale-invariant interactions with other particles 
are added, but, with the inclusion of scale-invariant gravitational 
interactions, $\abar,\bbar\ne0,$ that is in fact \emph{not\/} correct, since 
$\overline{\Delta\beta}_\xip \neq 0,$ \eqn{eq:deltabetaxipbar}.
 
In contrast, the gravitational contribution $\overline{\Delta\beta}_\xip$ to 
$\betabar_\xip$ vanishes for minimal coupling $(\xip = {1}/{6})$, 
$(\xi=0)$. In Einstein-Hilbert gravity, it is well-known that gravitons in curved 
spacetime are minimally coupled to scalars. This is another way in which 
gravitons differ from vector bosons, which are conformally coupled (in a 
renormalizable theory.) The beta-functions respect the symmetry properties 
operative at very short distances where IR irrelevant operators may be 
neglected. This property is quite general for perturbation theory in curved 
spacetime backgrounds; because of the equivalence principle, the local 
coupling of gravitons to scalars is as if spacetime were flat.

These observations can be made more quantitative by developing a systematic expansion in $\abar/\bbar$ near their UVFP. Given that 
$x=\bbar/\abar \gg 1$ near the UVFP, as a zeroth approximation, we may 
neglect the terms in $\abar$ in \eqn{eq:betabbar}, giving 
\beq\label{eq:betabbaraprx1}
\betabar_\bbar\approx\bbar\left[b_g-
\left(\frac{5}{12} +\frac{3 N(N{-}1)\xip^2}{4}\right)\bbar\right],
\eeq
As we shall confirm below, near the UVFP the $\xip^2$ term is completely 
negligible, so \eqn{eq:betabbaraprx1} is perfectly analogous to 
\eqn{eq:betaabar}, with the replacements 
$\{\abar\to\bbar, b_2\to \sfrac{5}{12}\}.$ Thus, in first approximation, 
$\bbar $ has a UVFP at $\bbar\uv\approx 12b_g/5,$ and 
$\betabar_\bbar\approx b_g(\bbar\uv-\!\bbar),$ which implies that, near their 
UVFPs, $\bbar\to\bbar\uv$ at the same rate as $\abar\to\abar\uv.$. As 
remarked earlier, typically, $12b_g/5\sim {\cal O}(1).$

In next approximation, suppose we neglect only the $\abar^2$ term on the 
right-hand side of \eqn{eq:betabbar}, then, neglecting the tiny term in 
$\xip^2,$ 
\eqn{eq:betabbaraprx1} would be replaced by 
\beq\label{eq:betabbaraprx2}
\betabar_\bbar \approx 
\bbar\left[b_g+ {\Large\sfrac{5\abar}{\bbar}} -
{\Large\sfrac{5\bbar}{12}}\right] \approx 
\bbar\left[\wt{b}_g-{\Large\sfrac{5\bbar}{12}}\right],\ \ \mathrm{where}\ \ 
\wt{b}_g\equiv b_g(1+5/b_2).
\eeq
In the second step, we replaced $\abar/\bbar$ by its asymptotic value and 
inserted our zeroth approximation for $\bbar\uv.$ Thus, 
$\bbar\uv\approx 12 \wt{b}_g/5.$
Although $1/b_2$ is relatively small, because it enters multiplied by 5, this 
correction can be important for obtaining an accurate estimate. For example, 
if $b_2=50 \gg 1,$ $5/b_2=1/10$, $\wt{b}_g/b_g=1.1,$ a 10\% increase over 
the zeroth approximation! To first order in $\abar/\bbar,$ we then 
get\footnote{This process could be iterated to further improve these estimates 
by including the $\abar^2$ term in $\betabar_\bbar,$ \eqn{eq:betabbar}, and 
expanding to higher orders in $\abar/\bbar$.} 
\beq\label{eq:xuvaprx}
\frac{\abar\uv}{\bbar\uv}=\frac{1}{x\uv}\approx
\frac{5 b_g}{12 b_2 \wt{b}_g}=
\frac{5}{12(b_2+5)}\ll 1,
\eeq
independent of $b_g$! E.g., if $b_2=50, x\uv\approx132,$ or 
${\abar\uv}/{\bbar\uv}\approx 0.76{\times}10^{-2}\ll 1.$

We can use these results to estimate $\xip\uv.$ From 
\eqns{eq:betaxipbar}{eq:deltabetaxipbar}, for conformal coupling
$(\xip = 0)$, we have
\beq\label{eq:betaxipbar0}
\betabar_\xip\Big|_{\xip=0}\hspace{-5pt}=
\overline{\Delta\beta}_\xip\Big|_{\xip=0}\hspace{-5pt}= 
-\frac{1}{6}\left[\frac{10 \abar}{3 x}\right]= 
-\frac{5 \abar^2}{ 9 \bbar}\approx 
-\frac{25 b_g}{108 b_2(b_2+5)}\sim
\ocal\!\left(b_g/b_2^2\right)\!,
\eeq
an extremely small number. For example, for $b_g=1, b_2=50,$ this gives 
${-}0.8{\times}10^{-4}.$
Since this is so small, it seems likely that $\xip\uv$ is nearby. 
In linear approximation,
\begin{subequations}
\begin{align}
\label{eq:betaxipbaraprx}
\betabar_\xip&\approx \overline{\Delta\beta}_\xip\big|_{\xip=0}+
\xip{\left[\betabar_\xip^{ \prime}\right]}_{\xip=0} ,\\
\label{eq:betaxipbarp}
\left[\betabar_\xip^{ \prime}\right]_{\xip=0}&
\approx\left[\frac{N^2{-}4}{24N} z_2+\frac{N(N{-}1)+ 4}{6}z_3-3(N{-}2)\right]+
\frac{3 \wt{b}_g}{5},\\
\label{eq:xipuv}
\xip\uv &\approx {-}\frac{\overline{\Delta\beta}_\xip}
{\left[\betabar_\xip^{ \prime}\right]}\Bigg|_{\xip=0},
\hspace{-8pt} \approx -
\frac{25 b_g}{108 b_2(b_2+5)} 
\bigg[\Big|\betabar_\xip^{ \prime}\Big|\bigg]_{\xip=0}^{-1}.\!
\end{align}
\label{eq:betaxipest}
\end{subequations}
\!\!(Here,\! ``betabar-prime" in $\betabar_\xip^{ \prime}$ denotes the 
partial derivative of $\betabar_\xip$ with respect to $\xip.$) 
These formulae require further explanation. From 
\eqn{eq:betaxipbar0}, we know that 
$\overline{\Delta\beta}_\xip|_{\xip=0}$ is very small and negative. 
Therefore, the linear approximation \eqn{eq:betaxipbaraprx} will yield a 
UVFP if and only if $[\betabar_\xip^{ \prime}]_{\xip=0}{<}0,$ which 
has been assumed in \eqn{eq:xipuv}. Once one obtains values for the 
UVFPs $\{z_2\uv, z_3\uv\},$ one must return to check this assumption, 
but it will be presumed to be true for the rest of this section. Then 
there is a UVFP at small negative $\xip$ given by \eqn{eq:xipuv}. The 
first contribution to the slope in \eqn{eq:betaxipbarp} comes from the 
first term in \eqn{eq:betaxipbar}, which arises from matter contributions 
in the absence of quantum gravity, i.e., QFT in curved spacetime. The 
second term in \eqn{eq:betaxipbarp} comes from the slope of 
$\overline{\Delta\beta}_\xip$, \eqn{eq:deltabetaxipbar}. Even though 
the second term in square-brackets in that formula vanishes at 
$\xip=0,$ it is the dominant contribution to the slope 
$\overline{\Delta\beta}^{ \prime}_\xip|_{\xip=0}=\bbar/4,$ the last 
term in \eqn{eq:betaxipbarp}, $\bbar\uv/4=3\,\wt{b}_g/5$. This is always 
positive and often not negligible. E.g., with $b_2=50,$ we have 
$\bbar\uv/4\approx 0.66\, b_g\sim\ocal(1).$

If the linear approximation breaks down, it is conceivable there could still be 
a UVFP of $\betabar_\xip$, but, for our $SO(10)$ model, \secn{sec:so10},
we numerically determined \emph{all\/} the FPs, which are listed in 
\tabbys{tab:FP1}{tab:FP2}, and there was no other UVFP. The linear 
approximation works extraordinarily well in this case; in \secn{sec:catch}, 
we provide a detailed comparison.

In general, to know the actual magnitude of $\xip\uv,$ \eqn{eq:xipuv}, 
we must know that there are UVFPs for $\{z_2,z_3\}$ and be 
able to at least estimate their values for input. To that end, we take up 
$\betabar{}_{z_2}, \betabar{}_{z_3},$ \eqns{eq:betaz2bar}{eq:betaz3bar}. 
Since the gravitational corrections 
$\overline{\Delta\beta}_1,\overline{\Delta\beta}_2,$ 
\eqns{eq:deltabeta1bar}{eq:deltabeta2bar}, do not depend explicitly on $\{z_2,z_3\},$ 
they may be estimated using the approximations in 
\eqns{eq:xuvaprx}{eq:xipuv}. Asymptotically, in each 
$\overline{\Delta\beta}_k$, we may replace $\{\abar, x, \xip \}$ by 
$\{\abar\uv, x\uv, \xip\uv \}.$ First consider $\overline{\Delta\beta}_1,$ 
\eqn{eq:deltabeta1bar}, which consists of two terms, the second of which is suppressed by $(x\xip)^2$ with respect to the first. From 
\eqns{eq:xuvaprx}{eq:xipuv}, we see that
\beq\label{eq:xxip}
x\xip \approx -\frac{5 b_g}{9 b_2} 
\Big[\big|\betabar_\xip^{ \prime}\big|\Big]_{\xip=0}^{-1} \ll 1.
\eeq
That being the case, certainly $(x\xip)^2$ is completely negligible with 
respect to the first term, so 
$\overline{\Delta\beta}_1\approx 5(\abar\uv/6)^2= 5(b_g/6b_2)^2 \lll 1.$ 
E.g., for $b_g=1,b_2=50,$ 
$\overline{\Delta\beta}_1\approx 0.6{\times}10^{-4}.$
Similarly, the second term in $\overline{\Delta\beta}_2$, 
\eqn{eq:deltabeta2bar}, is suppressed by $18(x\xip)\xip/5,$ also a 
negligible correction to the first term. Hence, 
$\overline{\Delta\beta}_2\approx 5 \abar\uv=5 b_g/b_2.$ 
For future reference, we note that both $\overline{\Delta\beta}_1$ and 
$\overline{\Delta\beta}_2$ are positive.

Before proceeding further with \eqns{eq:betaz2bar}{eq:betaz3bar}, we need to 
understand how roots of $\{\betabar{}_{z_2}, \betabar{}_{z_3}\}$ come about. 
We are only interested in models for which the UVFPs satisfy certain 
convergence criteria, \secn{sec:constraints}, and stability constraints, 
\secn{sec:ssb}. In the present context, the constraint of interest is that 
$\{z_2\uv, z_3\uv\}$ must both be positive. In that case, every term in 
$\betabar{}_{z_k}$ is positive except for the linear term 
${-}6(N{-}2)z_k,$~$(k{=}2,3).$ This sole negative term must cancel the 
sum of all the other terms\footnote{We use these observations 
in \secn{sec:catch} to set lower and upper bounds on the $z_k\!\uv$.}. 
It cannot be that each term becomes small, because, setting both 
$z_2$ and $z_3$ to zero, both $\betabar{}_{z_2}$ and 
$\betabar{}_{z_3}$ are large and positive.

Returning to our estimated gravitational corrections, we see from 
 \eqns{eq:betaz2bar}{eq:betaz3bar} that $\overline{\Delta\beta}_1$ contributes only 
to $\betabar{}_{z_3}.$ This is a very small positive constant to be added to 
the much larger one already present; with negligible error, 
we may drop $\overline{\Delta\beta}_1.$ Turning to 
$\overline{\Delta\beta}_2,$ we see that it enters both beta-functions in the 
coefficient of the terms linear in $z_k$ in the combination
$(b_g{+} \overline{\Delta\beta}_2) \approx \wt{b}_g,$ the same $\wt{b}_g$ that entered into the corrections to $\betabar_\bbar,$ \eqn{eq:betabbaraprx2}. 
Therefore, to a very good approximation sufficiently near the UVFP, we may replace \eqns{eq:betaz2bar}{eq:betaz3bar} with
\begin{subequations}
\label{eq:betazbar'}
\begin{align}
\label{eq:betaz2bbar'}
\betabar{}_{z_2}& = 36(N{-}8)+\frac{2N^2{-}N{-}24}{24N}z_2^2+ 4 z_3 z_2+
\left(\wt{b}_g-6(N{-}2) \right)z_2,\\
\label{eq:betaz3bbar'}
\begin{split}
\betabar{}_{z_3}& = \frac{36(N{-}2)}{N} +\frac{N(N{-}1){+} 16}{6} z_3^2
 + \frac{N^2{-}4}{48N^2} z_2^2+ \frac{N^2{-}4}{12N} z_3z_2+\hskip2mm\ \cr
&\hskip15mm
 \left(\wt{b}_g-6(N{-}2) \right)z_3.
\end{split}
\end{align}
\end{subequations}
These are identical to the flat-space beta-functions except for the 
replacement $b_g\to\wt{b}_g$! Since $\wt{b}_g{>}b_g{>}0,$ the effect of 
dynamical gravity is to increase the difficulty finding a UVFP of these two 
equations\footnote{Using the beta-functions of \reference{Salvio:2014soa}, 
we find the opposite sign of the effect reported in 
Refs.~\hspace{-2pt}\cite{Buchbinder:1989ma, Buchbinder:1989jd}.}. At least 
in the cases that we have examined, these equations are remarkably sensitive 
to the value of $\wt{b}_g,$ and we give an example in \secn{sec:catch}.

If there are real solutions for the roots of 
$\{\betabar{}_{z_2}, \betabar{}_{z_3}\}$, it remains to determine whether any 
of them is a UVFP by calculating the ``stability matrix" 
$\left[\pa\betabar{}_{z_j}/\pa{z_k}\right]$ at each FP and showing it has only 
negative eigenvalues. If such a UVFP candidate is identified, then one must 
return to \eqn{eq:betaxipbarp}, insert the values of $z_k$ at the FP, and check 
that $\overline{\Delta\beta}_\xip|_{\xip=0}{<}0,$ as we have assumed. 

We shall return to considering these equations for arbitrary $N$ 
elsewhere~\cite{einhorn:largeN}, but, in order to develop some intuition from 
experience with such models, we here restrict ourselves to $SO(10)$, which, 
for $N \geq 9$, is the smallest $SO(N)$ having complex spinor (i.e., chiral) 
representations. This is one reason why $SO(10)$ has been of great interest 
as a possible GUT.

\section{An \texorpdfstring{$SO(10)$}{so10} model}
\label{sec:so10}

Although our primary interest is in the existence of a UVFP in all the 
couplings, this model is simple enough to determine numerically 
\emph{all} the FPs of the exact one-loop beta-functions. Let us begin with 
$\betabar_\abar,$ 
\eqn{eq:betaabar}. 
As we discussed in the preceding section, $\abar$ has a
UVFP at $\abar\uv = b_g/b_2,$ whose dependence on $N$ is 
implicit through $b_g$ and $b_2$. 
For $N {=} 10,$ their values are
 $b_g = 4(21-T_F)/3, b_2 = (461+N_F)/20.$ 
Note that $T_F{<}21$ in order to preserve AF for the gauge coupling.
Setting $N {=} 10,$ the remaining beta-functions in \eqn{eq:betabars} are
\begin{subequations}
\label{eq:betabar10}
\begin{align}
\label{eq:betaxbar10}
\betabar_x& = 
\abar \left[-\frac{10}{3}+\frac{(6 N_F+3366)}{120} x-
\left(\frac{5}{12}+\frac{135}{2}\xip^2\right)x^2\right].\\
\label{eq:betaz2bar10}
\betabar{}_{z_2}& = 72+\frac{83}{120}z_2^2+4z_2z_3+\left(b_g-48\right)z_2+
\abar\left(5-18 x \xip^2\right)z_2,\\
\label{eq:betaz3bar10}
\begin{split}
\betabar{}_{z_3}& = \frac{144}{5}+\frac{53}{3}z_3^2+\frac{1}{50}z_2^2+\frac{4}{5}z_2z_3+ \\[-2mm]
&\hskip15mm (b_g-48)z_3+\abar\left(5-18 x \xip^2\right)z_3+\frac{\abar^2}{12}(6\xip-1)^2\left( 5+9x^2\xip^2 \right),
\end{split}\\
\label{eq:betaxipbar10}
\betabar_\xip& = \left(\frac{2}{5}z_2+\frac{47}{3}z_3-24\right)\xip+
\frac{\abar}{6}\left(6\xip-1\right)\left(\frac{10}{3x}-\frac{3}{2}x \xip(2\xip+1)\right).
\end{align}
\end{subequations}

Requiring that the gauge coupling be AF, $(b_g {>} 0),$ it would seem that
there are a large number of possibilities with $0 \leq T_F {<} 21$. In fact,
for reasons not particularly transparent, it turns out that there is a UVFP
only for $b_g$ as small as permitted. Restricting the fermions to be in the 
vector, spinor, or adjoint representations, $\{ {\bf 10}, {\bf 16}, {\bf 45} \}$, 
$T_F = 4n_1+\half n_2+n_3,$ and $N_F = 45 n_1+10 n_2+16 n_3,$ 
where $n_i$ is the number of representations (flavors) of each type. 
Since $b_g$ vanishes for $T_F = 21,$ the first allowable case has 
$T_F=\sfrac{41}{2}$ $(b_g=\sfrac{2}{3}).$ Even with $T_F$ fixed at 
$\sfrac{41}{2}$, there are still 66 possible choices for the three integers 
$(n_1,n_2,n_3),$ each with a different value for $N_F,$ spanning 
$235 \leq N_F \leq 410.$ This corresponds to the ranges 
$\sfrac{174}{5} \leq b_2 \leq \sfrac{871}{20},$ 
$0.015\lesssim\abar\uv\lesssim0.019.$ There is a UVFP for all 
values of $N_F$ in this range, and it is easy to see that the 
FPs are rather insensitive to $N_F.$ In \secn{sec:catch}, we show that, for 
$T_f=20,$ there is no UVFP.
\begin{table}[htbp]
\begin{center}
\begin{tabular}{|c|c|c|c|c| c| c| } \hline
& $ \abar $ & $ x $ & $ \xip $ & $ z_2$ & $z_3$ & Nature\\ \hline
$\!{\bf 1.} $ & $ {\bf 0.016856} $ & {\bf 106.8451} & ${\bf -1.4399 {\times}10^{-5}} $ & $ {\bf 1.7235} $
& $ {\bf 1.0706} $ & {\bf UV stable} \\ \hline
$2.\ $ & $ 0.016856 $ & $ 106.8450$ & $~1.0030 {\times}10^{-4}$ & $1.80221$ & $1.5129$ & saddle point\\ \hline
$3.\ $ & $ 0.016856 $ & $ ~0.07497$ & $ 0.10641$ & $ 1.80221 $ & $1.5130$ & saddle point\\ \hline
$4.\ $ & $ 0.016856 $ & $ ~0.07488$ & $\hskip-4mm -0.02161$ & $ 1.72354 $ & $ 1.0706 $ 
& saddle point \\ \hline
$5.^*\ $ & $ 0 $ & n.\ a. & $ 0 $ & $ 1.7180 $ & $ 1.0592 $ 
& saddle line \\ \hline
$6.^*\ $ & $ 0 $ & n.\ a. & $ 0 $ & $ 1.80134 $ & $ 1.5293 $ 
& saddle line \\ \hline
\end{tabular}
\caption{Fixed points for an $SO(10)$ model for finite $\abar$.}
\label{tab:FP1}
\end{center}
\end{table}

\noindent To illustrate, consider $(n_1,n_2,n_3) {=} (0,1, 20),$ for which
$N_F {=} 330.$ Then, $\abar\uv = \sfrac{40}{2373} \approx 0.016856.$ 
Inserting this value of $\abar$ into \eqn{eq:betabar10}, we find there
are still four FPs in the other coupling constants. In \tabby{tab:FP1},
we show the values we found for these 
four\footnote{*For lines $5.^*$ \& $6.^*,$
see Appendix~\ref{sec:IRFPs}.}${}^*$. To determine their
``nature", we must calculate the stability matrix by taking the
partial derivatives of the beta-functions with respect to each of the
variables, evaluating them at the FP, and determining the eigenvalues. 
As claimed, one of the FPs is UV stable. (For a model in which 3 (and
only 3) spinor representations do not acquire GUT-scale masses, the
alternative $(n_1,n_2,n_3) {=} (0,3,19)$, for example, might be preferable, 
with very similar results.) 

As expected from our discussion in the preceding section, at the UVFP, the 
value of $x\uv$ is large, while $\xip\uv$ is extremely small. We will defer to 
\secn{sec:catch} a more detailed quantitative accounting, but these 
approximations work extraordinarily well\footnote{For readers who wish to 
jump ahead, see the discussion surrounding \eqn{eq:betabbar10'/48}.}.

One may also explore whether there are FPs in the extreme IR limit. As 
mentioned earlier, the behavior of these equations in the IR limit is purely 
formal since, if weak coupling DT does not take place, then the gauge or 
gravitational interactions (or both) become strong, and perturbation theory 
breaks down. Nevertheless, understanding the IR behavior of the running 
couplings may help us more easily understand the range of couplings 
lying within the catchment basin of the UVFP. Since the determination of 
the IRFPs of these equations is not relevant to our main line of 
development, we have relegated that analysis to 
Appendix~\ref{sec:IRFPs}.

Having established the existence of a class of simple models 
with a UVFP, are there further restrictions on the allowed range of 
values of the coupling constants at the UVFP? In fact, as we 
shall discuss in the next section, there are.

\section{Constraints on the coupling constants}
\label{sec:constraints}

We have adopted the point of view of Euclidean quantum
gravity~\cite{Hawking:1978jn, Gibbons:1994cg}, in which the theory is
quantized starting from the Feynman path integral with Euclidean
signature, the Euclidean path integral (EPI) for short. Strictly
speaking, one must require this of the bare couplings defined in the
presence of a cutoff, and then show that one may obtain a sensible
renormalized theory as the cutoff is removed. As illustrated by the
enterprise of lattice field theory, this may be taken as a starting
point for a nonperturbative definition of a theory, but even so, it can
be problematic to remove the cutoff. For example, it is generally
believed that $\lambda\phi^4$ theory in four dimensions 
has no nontrivial continuum limit, the reason being that the 
renormalized interaction strength $\lambda$ at any finite 
scale tends to zero as the cutoff is removed. One case in which 
we can expect to find a continuum limit is in models in which 
all the couplings are AF. These are especially amenable to a 
perturbative treatment at high energies because
we are assured that the quantum corrections are small. This is
precisely the situation that has been established for the class of
theories under consideration here. 

The preceding considerations do not guarantee the existence of 
a sensible QFT. For example, $\lambda\phi^4$ in four dimensions
with $\lambda {<} 0$ is AF. We must require a convergent EPI 
at sufficiently high scales where the effective action may be 
approximated by the form of the ``classical" action 
with small couplings. Consider the action defined by 
\eqn{eq:hoaction} plus \eqn{eq:mataction} with the potential 
given in \eqn{eq:potvv}. We gather the result together here:
\beq
\label{eq:action4}
S_{cl}\! =\!\! \int\!\! d^4x\sqrt{g}\left[\frac{1}{4}\Tr[F_{\mu\nu}^2]\!+\!
\half \Tr[(D_\mu\Phi)^2]\!+\! \frac{h_3}{24} T_2^2\!+\!
\frac{h_2}{96} \widetilde{T}_4\!- \frac{\xi T_2R}{2}\!+\! 
\frac{C^2}{2a}\!+\! \frac{R^2}{3b}\!+\! c G\right]\!.\!
\eeq
It is not clear what constraint, if any, is implied by the presence of
the G-B term $G$. For now, we follow custom and ignore it, 
but we shall return to this question below. (It is certainly not ignorable 
in the determination of the effective action in de~Sitter space.)

Euclidean signature of the metric ensures that 
$\Tr[F_{\mu\nu}^2] \geq 0$ and $\Tr[(D_\mu\Phi)^2] \geq 0.$ For the 
integral over metrics at fixed other fields, the quadratic operators 
$C^2, R^2$ dominate for large fields. Therefore, both $a {>} 0$ and 
$b {>} 0$ since there are field configurations where one operator becomes 
large while the other does not. For the same reason, integration over the 
scalar fields implies both $h_2 \geq 0$ and $h_3 \geq 0.$ More 
generally, we must require that the quadratic form 
\beq
\label{eq:scalarconstraint}
\frac{R^2}{3b}-\frac{\xi T_2R}{2}+\frac{h_3}{24} T_2^2 \geq 0
\eeq
for all field configurations. Since $b{>}0,$ this form is positive as $T_2\to0,$ 
and it will have no real roots provided $h_3 \geq 9b\xi^2/2 {>} 0.$ 
This also implies that if either or both $\Phi$ and $R$ condense,
i.e., develop classical VEVs, then the associated cosmological constant
will be positive.

Altogether, we conclude that, at sufficiently large scales, we must have 
\begin{subequations}
\label{eq:constraints}
\begin{align}
&a>0,\quad b>0,\quad h_2>0,\quad 
h_3\geq \frac{9}{2} b \xi^2>0,\\
\label{eq:constraintsbar}
&\hskip-6mm \hbox{ or\ }\abar>0,\quad x>0,\quad 
z_2>0,\quad z_3\geq \frac{9}{2} x \abar \xi^2=\frac{9}{2} \bbar \xi^2>0,
\end{align}
\end{subequations}
where, in the second form, we have rewritten the constraints in terms
of the rescaled couplings after dividing by $\alpha \equiv g^2.$ From the
one-loop beta-functions, we know that, if $a {>} 0$ at some high scale,
then it will remain positive for all lower scales at which perturbation
theory remains valid. (Obviously, the same is true for $\alpha.$)
Note that the sign of $\xi$ is not constrained asymptotically, which
is fortunate since \eqn{eq:betaxipest} implied that the asymptotic value 
$\xip\uv\lesssim0,$ however tiny, so $\xi\uv\lesssim{-}1/6.$ (On the other hand, 
we demand $\xi {>} 0$ at the DT scale in order to induce normal Einstein gravity.)

Returning to the G-B term, $c G,$ one may write $G$ in the form
$\nabla_\mu B^\mu,$ where $B_\mu$ is a one-form, not a vector. 
In a smooth, compact background, the integral is proportional to the 
Euler number, which can take either sign, depending on the topology of 
the manifold. Thus, it is hard to imagine finding a constraint on the sign 
of $c.$ If one quantizes the theory using the background field method (BFM), 
then because $\sqrt{g} G$ has zero variation, it makes no contribution to the 
integral over the quantum fields. From this point of view, it is unnecessary to
constrain the coupling $c.$ Since the BFM is simply a change of field
variables, it ought to be true in general. It is not entirely clear that
this unambiguously defines the EPI nonperturbatively, but if we confine
ourselves to perturbation theory, then perhaps this argument is
sufficient to dispense with any constraints on the G-B coupling $c.$

Perturbatively, the value of $c$ is determined up to a constant $c_0$ by the other 
couplings in the theory. In fact, we showed~\cite{Einhorn:2014bka} that, 
in leading order, i.e., at tree level,
\beq
c = c_0-b_1/(b_2 a)=c_0-b_1/(b_2 \alpha \abar)
\eeq
 where the constants $b_1,b_2$ are given in \eqn{eq:bgb1b2}. 
Remarkably, at one-loop order, it, like $\beta_a$ and $\beta_\alpha,$ 
$c$ is independent of all the other couplings, including $b.$ 
Since $a$ and $\alpha$ go to zero asymptotically, clearly $c\to-\infty.$ 
(It appears as if $c_0$ would arise as a one-loop correction, but it is actually renormalization group invariant, i.e., it is scale independent.) Perhaps we 
should interpret AF to require $c_0=0,$ but it is not entirely clear how 
$c_0$ affects observables. In any case, it seems that it is not necessary to 
impose a constraint on $c$, but this may not be the final word on this subject.

\section{Spontaneous symmetry breaking}
\label{sec:ssb}

After this lengthy discussion concerning fixed points and UV behavior, 
we begin this section with an overview of the induced-gravity scenario 
that we have in mind. 
One solution of the classical field equations is 
the trivial solution $g_\mn {=} \eta_\mn, A_\mu {=} 0, \Phi {=} 0.$ One 
might think that this is the ``symmetric" phase in which none of the 
symmetries, including scale invariance, are broken, but, of course, 
scale invariance is explicitly broken in the QFT by the conformal 
anomaly, leading to a renormalizable theory rather than a conformal 
theory. Nevertheless, with all couplings AF, the theory does ultimately 
approximate a free field theory asymptotically, so this solution may be a 
possibility in the UV limit. For vector, scalar, and fermion fields, the 
elementary excitations are the familiar ones, but it isn't clear what 
particle-like excitations are to be associated with fluctuations in the 
metric, inasmuch as their propagators behave as $1/q^4.$ Despite 
that, this theory in the trivial background may be used to calculate 
beta-functions~\cite{Salvio:2014soa} and correlation functions at 
very short distances. There is nothing obviously problematic with this 
scalar-tensor theory so long as one realizes that it is limited in 
scope\footnote{We shall return to the question of whether this 
theory is unitary in \secn{sec:conclude}.}. 

The trivial solution is not however a solution for long-distances or 
low-energies, where, as we have described previously, there will be 
symmetry-breaking by DT, whether at weak or strong coupling. In order 
to realize something that looks more like our universe, it is crucial 
for consistency that scale invariance is anomalous and that the 
couplings run, so that we may entertain different approximate 
descriptions of the same underlying theory. (A strictly conformal 
theory with zero beta-functions is of little interest in this 
respect.) At a certain energy scale, set by DT, classical condensates 
form. If this occurs at weak coupling, as we assume in this paper, it 
is more nearly analogous to traditional GUT or electroweak symmetry 
breaking than to QCD: some of the massless particles simply acquire 
mass as a result of the formation of a scalar condensate, but also the 
curvature may become nonzero. Because the metric is associated with 
the geometry, the classical background may appear very different from 
Minkowski spacetime.

We can see how this works by reflecting on the form of the matter 
action, \eqn{eq:mataction}. From the scalar condensate, the 
coefficient of the scalar curvature becomes nonzero. This can be 
identified with the (reduced) Planck mass 
$\wt{M}_P^2\equiv\xi\vev{\Tr[\Phi^2]},$ where 
$\wt{M}_P\equiv M_P/\sqrt{8\pi}.$ At the same time, the condensate 
gives a nonzero value for the potential, which acts like a positive 
cosmological constant, $\Lambda\equiv\vev{V_J(\Phi)}/\wt{M}_P^2{>}0.$ 
Via the equations of motion, the curvature in first approximation has 
$\vev{R}=4\Lambda,$ as in Einstein-Hilbert theory. The simplest 
scenario would be a maximally symmetric background that approximates 
(half of) de~Sitter spacetime. It is not so clear what happens in a 
cosmological situation. It may be that the Big Bang begins when this 
condensate first forms, but we leave such questions for 
future research.

This is the induced-gravity mechanism; it is obviously generic, 
independent of the particular symmetry group or scalar content. 
One might think that it could not occur in perturbation theory, and, 
indeed, it may not. Spelling out the conditions under which that may 
occur is the subject of the remainder of this paper. 

The low-energy effective field theory, which includes a massless 
graviton in addition to massless matter, looks like ordinary general 
relativity plus matter. To leading order, of course, the graviton 
would appear to decouple, with interactions proportional to 
$1/\wt{M}_P.$ However, in de~Sitter background, there may be an 
exception to decoupling~\cite{Appelquist:1974tg}. With a nonzero 
cosmological constant, there remains an 
essential~\cite{Weinberg:1980gg} dimensionless coupling of the 
form $\Lambda/\wt{M}_P^2.$

The formation of a condensate $\vev{\Phi} \ne 0$ will also break the 
symmetry group $SO(10),$ and we shall see that the direction of the 
breaking \emph{can\/} be determined classically. Thus the mechanism 
that gives rise to the Planck mass and cosmological constant is also 
associated with the unification of gauge couplings. The particle physics 
will follow the familiar development, with some of the gauge bosons and 
scalars of $SO(10)$ acquiring masses, and others remaining massless. 
If fermions are added together with Yukawa couplings, some of them 
will also get masses. In the remainder of this section, we consider 
the classical breaking of $SO(10)$, which, it turns out, must be to 
$SU(5){\otimes}U(1)$. In the next section, \secn{sec:DT2}, we shall 
determine the DT scale, the energy at which these condensates form, 
while in \secn{sec:stable}, we shall investigate the stability 
requirements at the DT scale. 

Let us begin by analysing the extrema of the classical action to 
determine how $SO(10)$ might undergo spontaneous symmetry 
breaking (SSB). To solve the classical field equations in general is 
challenging when the background is curved and variable. To simplify 
the task, we shall assume that the background is approximately 
de~Sitter space and that any fluctuations in the curvature may be 
neglected in first approximation. (This is generally the case in 
inflationary models of the very early universe.) By dimensional analysis, 
$\int d^4x\sqrt{g} {=} V_4/R^2,$ where $V_4$ is an angular volume. 
In a de~Sitter-like background, the Weyl term contributes nothing,
but the Gauss-Bonnet (G-B) operator takes the value $G {=} R^2/6.$ 
Therefore, for constant $R$ and constant $\Phi,$ the value of the 
classical action takes the form
\beq\label{eq:action2}
\frac{S_{cl}}{V_4} = \frac{1}{3b}+\frac{c}{6}
+\frac{h_1}{24}\frac{T_2^2}{R^2}+\frac{h_2}{96} \frac{T_4}{R^2}-
\frac{\xi}{2}\frac{T_2}{R}.
\eeq
Since the action is dimensionless, it can depend only on
the ratio $\Phi/\sqrt{R}$, where we suppose that the 
relevant range of the scalar curvature has $\vev{R}>0.$ Classically, extremizing this action with
respect to $\Phi$ or $R$ will never yield a scale but it may fix their
ratio. The form of the action in \eqn{eq:mataction} has been treated 
many times, at least as far back as \reference{Li:1973mq}. One may 
employ the representation used therein, based on the standard form of 
the $SO(N)$ generators, or one may make a unitary transformation to bring the
generators to a form in which the Cartan subalgebra is represented by
diagonal matrices. (See, e.g., \reference{Cahn:1985wk}.) The latter 
are particularly simple. The generators take the form 
\beq\label{eq:basis}
R^a=
\begin{pmatrix}
\ \rcal_1 &\vline& \ \rcal_2 \\
\hline
- \rcal_2^* &\vline& - \rcal_1^t
\end{pmatrix},
\eeq
where the $ \rcal_i$ are $5{\times}5$ matrices with the properties
$\rcal_1$ is Hermitian and $ \rcal_2$ is antisymmetric. (Here, $ \rcal_1^t $ 
denotes the transpose.) We shall regard the elements of $R_1$ and the 
nonzero elements of $R_2$ as our $25{+}20{=} 45$ independent dynamical 
real variables. Defining $\varphi\equiv \Phi/\sqrt{R},$ the first variation of the 
action \eqn{eq:action2} is\footnote{$\delvar$ is shorthand for a matrix of the 
form of \eqn{eq:basis} with Hermitian $\delvar_1$ and antisymmetric 
$\delvar_2$.} 
\beq\label{eq:var1}
\frac{\delta S_{cl}}{V_4} = \frac{h_1t_2}{6}\Tr[\varphi\delvar]
+\frac{h_2}{24} \Tr[\varphi^3\delvar]-\xi\Tr[\varphi\delvar],
\eeq
where $t_2{\equiv}\Tr[\varphi^2].$ 
The vanishing of this equation for arbitrary $\delvar$ determines the extrema 
$\!\vev{\varphi}$. 

Assuming that $\!\vev{\varphi}$ is constant and nonzero, one 
may apply an $SO(10)$ transformation to bring $\!\vev{\varphi}$ into 
diagonal form. Calling the five real entries in 
$ \vev{\varphi_1} \equiv \mathrm{Diag}(r_1,r_2,r_3,r_4,r_5),$ 
then $\vev{t_2} = 2\sum_1^5 r_i^2$, and the vanishing of \eqn{eq:var1} takes 
the form
\beq
\label{eq:extrema1}
\left(\frac{h_1\vev{t_2}}{3}-2\xi\right)\Tr[ \vev{\varphi_1}\delvar_1]
+\frac{h_2}{12} \Tr[\vev{\varphi_1^3}\delvar_1]=0.
\eeq
Clearly, only the diagonal elements of $\delvar_1$ enter this equation; since 
they are independent, this implies 
\beq
\label{eq:extrema2}
r_j\left[ \frac{h_1\vev{t_2}}{3}-2\xi+\frac{h_2}{12}r_j^2 \right]=0,
\eeq
for each element $r_j, j=\{1,\ldots,5\}.$ Consequently, $\!\vev{\varphi_1}$ has 
diagonal entries either $r_j {=} 0$ or $r_j {\equiv} \pm r_0^{[k]},$ 
with $r_0^{[k]}$ satisfying
\beq
\label{eq:classicalssb}
\frac{h_1\vev{t_2}}{3}-2\xi+\frac{h_2}{12}r_0^{[k]}{}^2=0.
\eeq
Here, $k$ denotes the number of zero elements along the diagonal
$k = \{0,\ldots,4\}$. All nonzero elements have the same magnitude,
$r_0^{[k]},$ so there are five possible nontrivial extrema with
$r_j = r_0^{[k]}\omega_k$ 
with\footnote{In fact, any of the
nonzero entries could be $-1$ instead, but this is not really distinct. 
WLOG, one may exchange the negative entry in $ \rcal_1$ with the
corresponding positive element in $- \rcal_1^t$.}
$\omega_0 \equiv \mathrm{Diag}(1,1,1,1,1)$,
$\omega_1 = \mathrm{Diag}(1,1,1,1,0)$, $\ldots,$ 
$\omega_4 = \mathrm{Diag}(1,0,0,0,0).$ Correspondingly, 
$\vev{t_2} = 2(5-k)r_0^{[k]}{}^2,$ so
\beq
\label{eq:r0k}
r_0^{[k]} = \sqrt{\frac{24\xi}{8(5-k)h_1+h_2}}.
\eeq
As already remarked in section~\ref{sec:constraints}, we require 
$\xi {>} 0$ at the DT scale in order to generate a ``right sign'' Einstein 
term; moreover, as we shall see shortly, we must in any event have 
$\xi {>} 0$ at the DT scale for classical stability of the symmetry breaking. 
So we must also require that $h_2{+}8(5-k)h_1 {>} 0$ in order to have 
real solutions for $r_0^{[k]}.$ We previously argued that, for the EPI to 
converge, \eqn{eq:constraints}, we must have $h_2 {>} 0$ and 
$h_3 {=} h_1{+}h_2/40 {>} 0$ {\it asymptotically\/}, but these constraints are 
not necessarily true at the DT scale. In fact, however, we shall see below 
that in this simple model, stability of the $SO(10)$ breaking requires the number of zero elements $k {=} 0,$ with $(\xi,h_2,h_3) {>} 0.$

To explore local stability of these five extrema, we must determine the 
second variation of the action. Returning to \eqn{eq:var1}, the second 
variation is 
\begin{align}\label{eq:var2}
\begin{split}
\frac{\delta^2 S_{cl}}{V_4}&=\left\{\frac{h_1}{3}(\Tr[\varphi\delvar])^2+\frac{h_2}{24}\left[2\Tr[\varphi^2\delvar^2]+\Tr[(\varphi\delvar)^2] \right]
\right\}\\
&+\left(\frac{h_1 t_2}{6}-\xi\right)\Tr[\delvar^2].
\end{split}
\end{align}
To determine whether the candidate vacua are stable, we must evaluate 
\eqn{eq:var2} for $\varphi \to \vev{\varphi^{[k]}}$ and arbitrary $\delvar$.
This is a rather complicated equation involving four distinct traces. We shall 
simply state the result here and refer the interested reader to 
Appendix~\ref{sec:var2} for details. We find that the only local minimum 
among the five extrema has the number of zeros $k{=}0$, provided that 
$\{\xi,h_2,h_3\}$ are all positive\footnote{After including radiative corrections, 
these turn out to be necessary but not sufficient conditions, as we shall 
discuss in \secn{sec:stable}.}. Thus, we have classical stability at the DT 
scale only for breaking to $SU(5){\otimes}U(1)$. It is interesting that 
this specific breaking pattern is singled out in this approach and preferred to 
other popular alternatives, such as $SU(4){\otimes}SU(2){\otimes}SU(2)$. 


 Moreover, the maximal subgroup $SU(5){\otimes}U(1)$ of $SO(10)$ is 
precisely the group associated with ``flipped'' $SU(5)$ 
models~\cite{Barr:1981qv}. (Of course, in the absence of fermions, 
we do not distinguish this possibility from Georgi-Glashow 
$SU(5){\otimes}U(1)$. For a recent analysis of ``flipped" phenomenology, 
see, for example, \reference{Ellis:2011es}.)

As remarked in the preceding section, asymptotically, we also 
must have $(h_2 {>}0,$ $h_3 {>}0)$ for convergence of the EPI. In fact, the 
UVFP in \tabby{tab:FP1} fulfilled these conditions but has $\xi {<} 0,$ 
so that the sign of $\xi$ must change while running from the DT scale
(where we require $\xi {>} 0$) to its UVFP. This turns out to be possible. 

Even though we have determined the symmetry-breaking pattern, the 
actual value of the DT scale remains to be determined. We want to 
show that the RG evolution fixes the DT scale while allowing for all 
these stability conditions to be fulfilled. This is the topic to which 
we shall turn in the next section.

Before so doing, a final remark: for this particular symmetry breaking, the 
coupling constant $h_3$ is to be preferred to $h_1$, which is reinforced by 
noting that the value of the classical action on-shell after symmetry breaking is 
given by
\beq
\label{eq:valaction}
\frac{S_{cl}^{(os)}}{V_4}=\frac{1}{3b}+\frac{c}{6}-\frac{3\xi^2}{2 h_3}.
\eeq
This is because $\widetilde{T}_4 {=} 0$ for this breaking pattern, and 
$\vev\Phi$ is $SU(5){\otimes}U(1)$ invariant.



\section{Dimensional Transmutation}
\label{sec:DT2}

In our paper on scale invariance~\cite{Einhorn:2014gfa}, we derived the
conditions for DT in models like this one. 
The effective action takes the generic form 
\beq
\label{eq:effaction}
\Gamma(\lambda_i,r,\rho/\mu) = S_{cl}(\lambda_i,r) 
+ B(\lambda_i,r)\log(\rho/\mu) + \frac{C(\lambda_i,r)}{2}\log^2(\rho/\mu) 
+\ldots,
\eeq
where $\rho \equiv \sqrt{R}.$ All coupling constants are denoted by the
set $\{\lambda_i\}.$ In writing the effective action in this
form, we have assumed that $\Phi$ is spacetime independent and 
that the background metric is well-approximated by
the de~Sitter metric with constant scalar curvature $R.$ (In general, we would have to return to the Lagrangian
form analogous to \eqn{eq:action4} rather than to this integrated action
analogous to \eqns{eq:action2}{eq:valaction}.) 
The functions
$B(\lambda_i,r),$ $C(\lambda_i,r)$ remain to be determined. In the
loop-expansion, $B {=} B_1+B_2+\ldots,$ with the first nonzero contributions
to $B$ coming at one-loop. Similarly, $C {=} C_2+C_3+\ldots,$ with the
first nonzero contributions to $C$ starting at two-loops. 

In this section, we shall evaluate $B_1^{(os)}$ and, in the next
section, $C_2^{(os)};$ here $``(os)"$ signifies ``on-shell'', that is to
 say evaluated with $r {=} r_0^{[0]}$ and $\mu {=} \vev\rho {=} v$. The 
 classical action $S_{cl}(\lambda_i,r)$ plays a central role in these 
calculations, so we begin by reviewing some of its features in our $SO(10)$ 
model. We shall need it off-shell in the next section, for which the general form 
was given in \eqn{eq:action2}, with the first and second variations in 
\eqns{eq:var1}{eq:var2}.

For our purposes in this section, we may assume the breaking is in the 
$SU(5){\otimes}U(1)$ direction, so that $r_i^2 \equiv r^2$ for all $i$. 
Then $S_{cl}$ becomes
\beq
\label{eq:scloffshell}
\frac{S_{cl}}{V_4}=\frac{1}{3b}+\frac{c}{6}+\left(\frac{25 h_3}{6} \right)r^4-5\xi r^2,
\eeq
where $r \equiv \sqrt{T_2/(10R)}$. Although we specified the direction
of the breaking, we have not put the ratio $r$ on-shell. The first
and second derivatives of this expression are 
\beq
\frac{S'_{cl}}{V_4}=10 r\left[\frac{5}{3}h_3 r^2-\xi \right], \quad
\frac{S''_{cl}}{V_4}=10\left[5h_3 r^2-\xi\right].
\eeq
As was previously noted toward the end of \secn{sec:ssb},
the first derivative vanishes for $r \to r_i^{[0]} {=} \sqrt{3\xi/(5h_3)},$ 
where the classical curvature becomes
\beq
\label{eq:clcurv}
\frac{ S''_{cl} }{V_4}\Big|_{r_i}=20 \xi.
\eeq
We see that, in order that the ratio of fields 
$\vev{T_2(\Phi)}\!/\!\vev{R}$ be classically stable, 
we must have $\xi {>} 0$. 

The value of the scalar curvature $\vev{R}$ is undetermined
classically, and the normalization scale of the couplings 
$h_3(\mu),\xi(\mu)$ is also unknown. 
We want to determine where the first
derivative of the effective action \eqn{eq:effaction} with respect to
$\rho$ vanishes. Taking the couplings to be normalized at the scale
of the breaking where 
$\rho\equiv v,$ then the extrema are determined at one-loop order by the 
vanishing of $B_1$ on-shell, where it takes the generic form\footnote{Since 
the classical action and beta-functions are real, if the effective action had an 
imaginary part, this procedure would not find it. One would have to return 
to calculating the radiative corrections directly.}
\beq
\label{eq:B1os}
B_1^{(os)}(\lambda_i(v),r_0)=
\sum_i\beta_{\lambda_i(v)}\frac{\pa S_{cl}}{\pa\lambda_i}
\Big|_{r=r_0}\!=0,
\eeq
which is to be evaluated at its extremum (either before, as 
in \eqn{eq:valaction}, or after taking its derivatives.) 
In \eqn{eq:B1os}, the quantity $r_0^{[0]}{}^2,$ \eqn{eq:r0k}, 
has been replaced by the rescaled ratio $r_0^2 \equiv 3\xi/(5z_3)$ 
to make manifest that $B_1^{(os)}$ is a function of the ratios only!

Actually, we can pause here to ask whether the Yang-Mills $SO(N)$ without 
any other form of matter can undergo DT. The only couplings would then be 
$\{a,b,\alpha\},$ and $B_1^{(os)}$ is a function of the two ratios $\{\abar,\bbar\}$ 
only. This calculation is quite similar to the one carried out for pure gravity 
earlier in \reference{Einhorn:2014gfa}. The qualitative results are the same, viz., one can 
in fact satisfy the $B_1^{(os)}{=}0$ for a certain value of $w{=}\abar/\bbar,$ but it is always locally unstable ($C_2{<}0$ in the language of 
\secn{sec:stable}.) This remains true if one adds an arbitrary number of 
fermions. We shall not stop to discuss this calculation.

Returning to \eqn{eq:B1os} for the present model, inserting the one-loop 
beta-functions for the couplings, and rewriting everything in terms of the 
rescaled variables $\{\abar,\bbar,z_2,z_3\}$ defined in \eqn{eq:rescaled}, 
we find
\beq
\label{eq:B1onshell}
B_1^{(os)}(z_3, z_2, \xip, x, \abar)
=\frac{b_3(x,\xip)}{3x^2}-\frac{b_1}{6}
-\frac{25 r_0^4}{6}\left( \betabar{}_{z_3}-b_g z_3\right)-5 r_0^2 
\betabar_\xip,
\eeq
where $b_1,b_g$ may be found in \eqn{eq:betagac}; $b_3,$ in 
\eqn{eq:betab}. $b_3$ is essentially the beta-function for $b$ and is 
closely related to $\betabar_x,$ as can be seen in 
\eqn{eq:betaxbar}. Note that the G-B beta-function $b_1$ 
contributes in an important way. 

In our $SO(10)$ model, with a single real adjoint scalar and $T_F{=}41/2,$ the parameters take the values 
\beq
\label{eq:parameters}
b_g{=}\frac{2}{3},\ b_1{=}\frac{(8806+N_F)}{720},\ b_2{=}\frac{(461+N_F)}{20},\ 
b_3{=}\frac{10}{3}-5 x+\left(\frac{5}{12}+\frac{135 \xip^2}{2}\right)x^2\!,
\eeq
and $\betabar{}_{z_3}$ and $\betabar_\xip$ may be 
taken from \eqns{eq:betaz3bar10}{eq:betaxipbar10}, respectively.
Thus, $B_1^{(os)}$ is independent of
$\alpha$ and depends only on the ratios of couplings via the various $
\betabar $'s. The absolute magnitudes of the couplings 
$\{\alpha, a, h_2, h_3\}$ are irrelevant so long as they are within the
perturbative regime. The explicit form of \eqn{eq:B1onshell} is long
and complicated; it is given in \eqn{eq:b1osfull} of 
Appendix~\ref{sec:dtscale}.

There are also some constraints that we must apply from our discussion 
of SSB in \secn{sec:ssb}: In order for SSB of $SO(10)$ to occur, we 
found that $h_3 {>} 0,$ and, for local stability of that breaking pattern,
$h_2 {>} 0.$

\begin{figure}[htbp]
\centering
\setlength\fboxsep{4pt}
\fbox{\includegraphics[width=4.in]{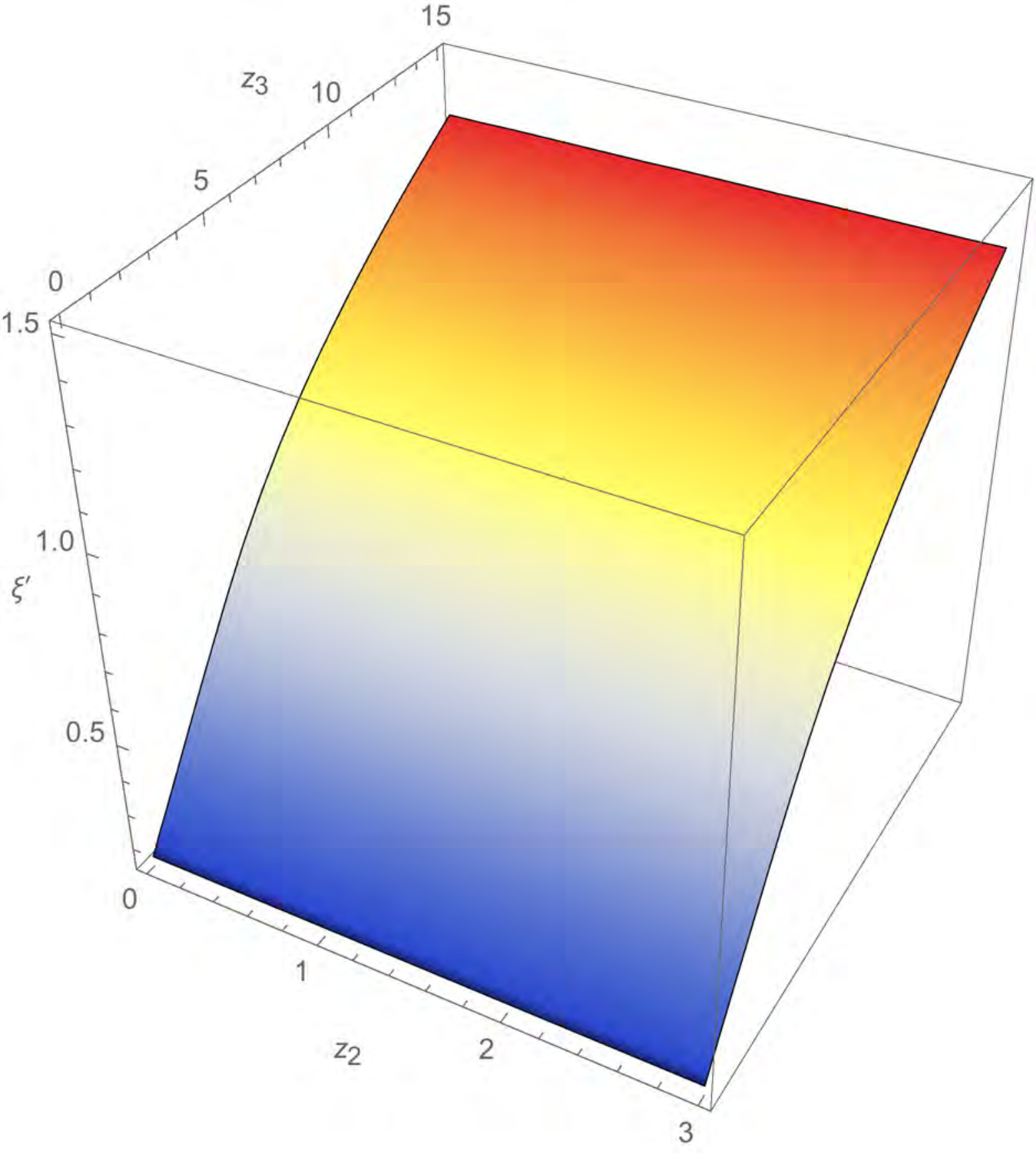}}
\caption{Portion of DT-surface for $x=120, \abar=0.025$}
as function of $\{z_2, z_3, \xip\}$ 
\label{fig:dtsurface1}
\end{figure}

In sum, in addition to $B_1^{(os)} {=} 0$ at the DT scale, we require 
$\{ \abar, \xi, z_2, z_3 \}$ positive, $(\xip{>}1/6).$ We refer to the 
range of couplings satisfying all these requirements\footnote{We do not 
include the constraints of stability under radiative corrections 
(\secn{sec:stable}) or lying in the catchment basis of the UVFP 
(\secn{sec:catch}).} as the {\bf DT-surface} in the five-dimensional space 
$\{ \abar, x, \xip, z_3, z_2 \}.$ In \figr{fig:dtsurface1}, we display a small 
portion of the DT-surface in the case presented in 
\tabby{tab:FP1}, viz., in which the fermion content corresponds to 
$T_F\, {=} 41/2, N_F {=} 330.$ In the figure, we chose to portray a 
three-dimensional slice of the DT-surface having $x {=} 120, \abar {=} 0.025.$ 
(There is a continuum of other slices possible.)

There are two more crucial restrictions on the portions of the 
DT-surface that are acceptable candidates for symmetry breaking. 
The first is to determine the nature of the stationary point at $\rho {=} v$
and to find that subregion of the DT-surface for which this is a local
\emph{minimum} of the effective action\footnote{It may be sufficient
to be metastable, if the lifetime is longer than the age of the
universe, but we would expect this to be only be a slight extension of
the stable region.}, a requirement equivalent to requiring the dilaton
$(\hbox{mass})^2$ to be positive. As with the running of the couplings, this
nonzero mass is due to the conformal anomaly, but, unlike the DT 
scale, the leading contributions to it are two-loop order. In the next 
section, we shall determine this eigenvalue of the second variation of 
the action called $\varpi_2.$
The second restriction is the nontrivial requirement that the couplings
lie within the basin of attraction (or catchment) of the UVFP. (This is
where our previous attempts~\cite{Einhorn:2015lzy} failed.) We take
this up in \secn{sec:catch}.
Both of these additional restrictions are complicated, 
the first, because it occurs at two-loop order, and the second, 
because it involves the full nonlinearities of the beta-functions.

\section{Local Stability of the DT-surface}
\label{sec:stable}

Our goal in this section is to determine the conditions under which
portions of the DT-surface are locally stable. The effective action has
the generic form~\cite{Einhorn:2014gfa} given in
\eqn{eq:effaction}. We shall replace $B$ by $B_1$ and $C$ by 
$C_2$, their leading non-zero contributions. Using the Renormalisation 
Group Equation for $\Gamma$ as defined in \eqn{eq:effaction}, we 
found that, off-shell, $B_1(\lambda_i,r)$ and $C_2((\lambda_i,r)$ 
satisfy:
\begin{subequations}
\label{eq:oneloopoffshell}
\begin{align}
B_1(\lambda_i,r)&=
\beta_{\lambda_i}^{(1)}\frac{\partial}{\partial \lambda_i }S_{cl}(\lambda_i,r)-
\gamma_r^{(1)} r S'_{cl}(\lambda_i,r), \\
B'_1(\lambda_i,r)&=\beta_{\lambda_i}^{(1)}\frac{\partial}{\partial \lambda_i }
S'_{cl}(\lambda_i,r)-
\gamma_r^{(1)}\frac{\partial}{\partial r}\big(rS'_{cl}(\lambda_i,r)\big), \\
C_2(\lambda_i,r)&=
\left[\beta_{\lambda_i}^{(1)}\frac{\partial}{\partial \lambda_i }
-\gamma_r^{(1)} r\frac{\partial}{\partial r}\right]B_1(\lambda_i,r),
\end{align}
\end{subequations}
where $\gamma_r^{(1)}$ is the one-loop anomalous dimension of the 
field, and we have suppressed other possible gauge-dependent terms 
that would contribute off-shell in gauges in which the RGE contains a 
gauge parameter. These equations are quite general and, in particular, do not 
require the classical action to be broken in the $SU(5){\otimes}U(1)$ direction. 

In our earlier paper~\cite{Einhorn:2014gfa}, we showed that the second 
variation of the effective action, \eqn{eq:effaction}, is given on-shell by
\beq
\label{eq:deta2onshell}
\delta^{(2)}\Gamma=\frac{1}{2}
\begin{pmatrix}
 \delta r & \frac{\delta\rho}{\rho}
\end{pmatrix}
\begin{bmatrix}
 S_m^{\prime\prime}(\lambda_i,\!r_0) &\ B_1{}^\prime(\lambda_i,\!r_0)\\
B_1{}^\prime(\lambda_i,\!r_0) &\ C_2(\lambda_i,\!r_0)
\end{bmatrix}
\begin{pmatrix}
 \delta r \\ \frac{\delta\rho}{\rho}
\end{pmatrix}.
\eeq
This is a kind-of see-saw situation, since $S_m^{\prime\prime}$ is $O(1)$; $B'_1$, 
$O(\hbar)$; to $C_2$, $O(\hbar^2).$
This stability matrix has eigenvalues equal to 
$S^{\prime\prime}_{cl}(\lambda_i)/2+O(\hbar),$ and 
\beq
\label{eq:eigenvalues}
\varpi_2(r_0,v)=\half\left[C_2-\frac{\left(B'_1\right)^2}{S_{cl}^{\prime\prime}}\right]_{r=r_0}+O(\hbar^3).
\eeq
In the case of breaking to $SU(5){\otimes} U(1)$, 
 we have from \eqn{eq:clcurv}, $S''_{cl}/2 {=} 10\xi,$ but we need to 
calculate the corresponding $\varpi_2(r_0,v)$ for our present theory.
Although of two-loop order, $C_2$ is evidently computable from
\eqn{eq:oneloopoffshell} knowing only the one-loop results. Since the
anomalous dimension $\gamma_r^{(1)}$ cancels out on-shell in 
$\varpi_2(r_0,v)$, we shall ignore it in the following and simply
compute the terms we need to determine $\varpi_2.$ In
\eqn{eq:B1onshell}, we only gave the form of $B_1$ on-shell, but here we
need it off-shell in order to determine $B'_1.$ In fact, the terms
have essentially the same form as before with the replacement of $r_0$
by $r$. Then we can compute
\begin{subequations}
\begin{align}
B_1
&=-\frac{\beta_b}{3b^2}-\frac{b_1}{6}+
\frac{25 r^4}{6}\left(\betabar{}_{z_3}-b_g z_3\right)-5 r^2 
\betabar_{\xi}+\ldots,\\
B'_1&=\frac{50 r^3}{3}\left(\betabar{}_{z_3}-b_g z_3\right)-
10 r \betabar_{\xi} +\ldots\\
C_2&=\beta_{\lambda_i}^{(1)}
\frac{\partial}{\partial \lambda_i }B_1(\lambda_i,r)+\ldots,
\end{align}
\end{subequations}
where the ellipses represent gauge-dependent terms that are essentially
irrelevant in that they cancel out in $\varpi_2$. The actual analytic
expression is reproduced in \eqn{eq:varpi2osfull} of 
Appendix~\ref{sec:dtscale}. The requirement that $\varpi_2(r_0,v) {>} 0$
turns out to be a strong restriction on the portions of the
DT-surface that are allowed. 

\begin{figure}[thbp]
\centering
\setlength\fboxsep{4pt}
\fbox{\includegraphics[width=4in]{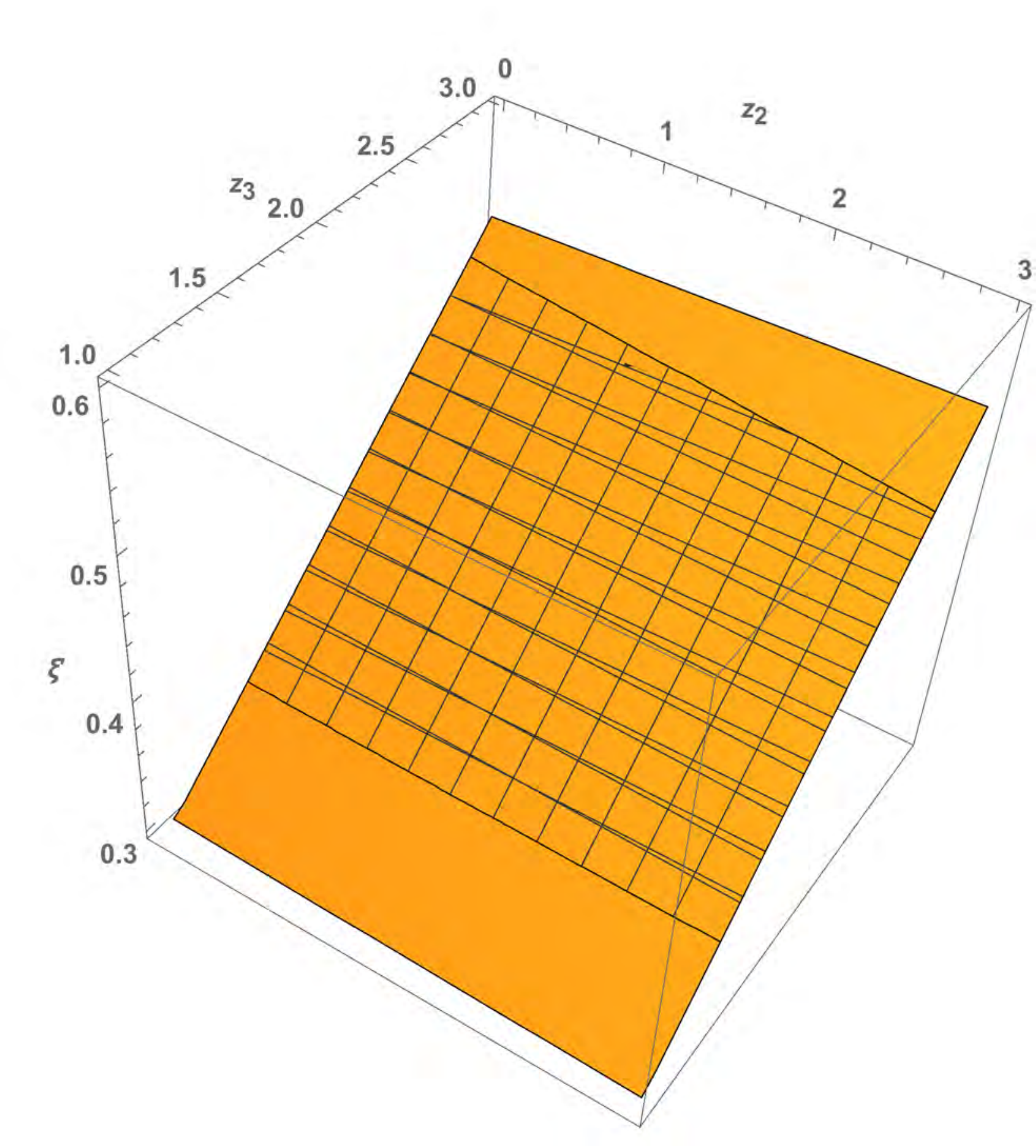}}
\caption{Section of DT-surface having $\varpi_2{>}0.$}
Cross-hatched portion has $\varpi_2{>}0.$
(Same parameters as in \figr{fig:dtsurface1})
\label{fig:dtsurface2}
\end{figure}

As an illustration, in \figr{fig:dtsurface2}, we display a subsection of the 
DT-surface shown in \figr{fig:dtsurface1}, with the same parameters as 
given there. The impact of the restriction to $\varpi_2 {>} 0$ is shown 
by the cross-hatched region.

Of course, the dilaton $(\hbox{mass})^2$ is proportional to\footnote{The
exact relation depends on the normalisation of the $\rho, \Phi$
kinetic terms. It is most simply and reliably determined in the Einstein
frame and will be spelled out in a future 
publication~\cite{Einhorn:tbp}.}  $\varpi_2(r_0,v)v^2,$ so
local stability is equivalent to requiring that the dilaton is not
tachyonic. The gauge bosons of $SU(10)/SU(5){\otimes}U(1)$ 
obtain masses of $O(g_{DT}\vev\Phi)$, where $g_{DT}$ is the gauge
coupling at the DT scale (the gauge unification scale), 
and $\vev\Phi \sim v\sqrt{\xi/h_3}$.

The requirement $\varpi_2(r_0,v) {>} 0$ completes the set of relations 
on the ratios of coupling constants\footnote{Rather than repeat long 
phrases such as this one or ``coupling constant ratios," we shall 
refer to them as ``couplings" or ``ratios" when it should be clear 
from the context what is intended.} that must obtain on the DT-surface. 
However, we must also know which points in this subregion actually lie 
within the basin of attraction of the UVFP. 

\section{The Catchment Basin of the UVFP}
\label{sec:catch}

In the preceding sections, we have specified all the requirements for 
the existence of a DT scale where symmetry-breaking occurs in a manner 
that is locally stable. To review, we seek points on the DT-surface 
that, for classical stability, have $\{ \abar, \xi, z_2 , z_3 \},$ all positive; 
in addition, for stability under quantum fluctuations, 
$\varpi_2(r_0,v) {>} 0.$ All of these 
conditions can be expressed in terms of these five ratios, but 
we tacitly assume that the original six couplings, 
$\{ \alpha, a, b, \xi, h_2, h_3 \},$ were small enough to justify the 
use of perturbation theory. Of course, if the five ratios are all less 
than one at their UVFP, then in the absence of data to the contrary, 
we may simply choose $\alpha(v)$ to be small at the scale 
$v.$ Possibly relevant data comes from searches for proton 
decay\footnote{For a review, see S. Raby, 
``Grand Unified Theories" in \reference{Agashe:2014kda}.} that place 
the scale of gauge coupling unification around $10^{16}$~GeV, 
where $SU(5){\otimes}U(1)$ may have been broken, so the unification 
to $SO(10)$ is at least that large. An estimate of the gauge coupling 
at that scale is $g^2/(4\pi)\approx 0.04,$ or $g^2\approx 0.5.$

Once one has a set of couplings ratios $\{ \abar, x, \xip, z_3, z_2 \}$ 
 fulfilling all the preceding conditions on the DT-surface, one must 
ascertain whether or not a given point flows to the UVFP so that the 
running couplings are AF. This is by no means trivial; often one or
another of these ratios blows up rather than approaching the UVFP. With
reference to \tabby{tab:FP1}, we see, for example, that the saddle
point on line~2 lies very near the UVFP. A saddle repels couplings
coming from one direction while attracting them from another. Thus, a
linearized analysis is of little use over a large range of scales, and
there is no alternative to starting at a point on the DT-surface that
also is locally stable and running the couplings up to higher scales
in order to determine whether the five coupling constant ratios
approach their UVFP. This is exactly what is done in the SM from the
electroweak scale to test for gauge coupling unification. Here we must
test whether they flow to the UVFP or lead to a breakdown of
perturbation theory. 

We recall that the equations that must be solved for the running 
couplings take the form
\beq
16\pi^2\frac{d \lambda_i}{du}=\betabar_{\lambda_i},
\eeq
where $du \equiv \alpha(t) dt.$ Here, $\lambda_i$ represents any of the five ratios of coupling constants $\{ \abar, x, \xip, z_3, z_2 \}.$ The corresponding $\betabar_{\lambda_i}$ 
are given in \eqns{eq:betaabar}{eq:betabar10}. 
We may infer certain general properties from the form of these beta-functions. 
The couplings $a(\mu)$ and $\alpha(\mu)$ do not mix with other couplings at 
one-loop, so being positive asymptotically, they remain so as the scale $\mu$ 
decreases. Consequently, their ratio $\abar$ also remains positive at the DT 
scale. As discussed earlier in \secn{sec:betas}, 
$\betabar_\abar,$ \eqn{eq:betaabar} has its UVFP at $\abar\uv {=} b_g/b_2.$ 
All $\abar(\mu){>}0$ flow monotonically to this UVFP, so long as the initial 
values of $a$ and $\alpha$ lie within the perturbative domain. As 
$\abar \to \abar\uv,$ $\betabar_\abar \approx b_g(\abar\uv-\abar),$ 
so its final rate of approach is set by $b_g.$

\begin{figure}[htbp]
\centering
\begin{subfigure}[htp]{.5\textwidth}
\centering
\setlength\fboxsep{4pt}
\fbox{\includegraphics[width=\textwidth]{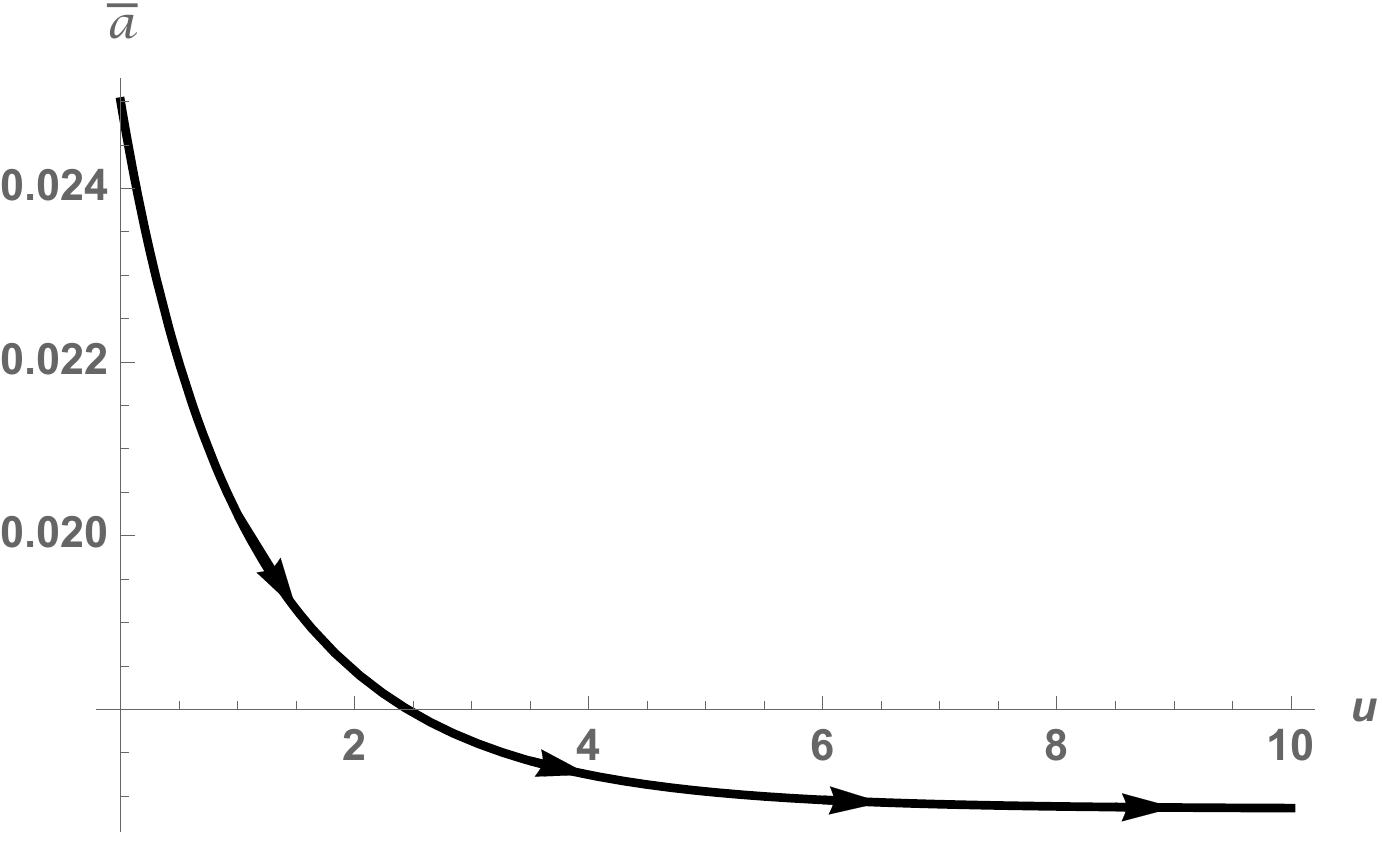}}
\caption{ $\abar$ from $0.025 \to 0.016856.$} 
\label{fig:runningabar}
\end{subfigure}
\linebreak
\begin{subfigure}[htp]{.47\textwidth}
\hskip-.2in
\centering
\setlength\fboxsep{4pt}
\fbox{\includegraphics[width=\textwidth]{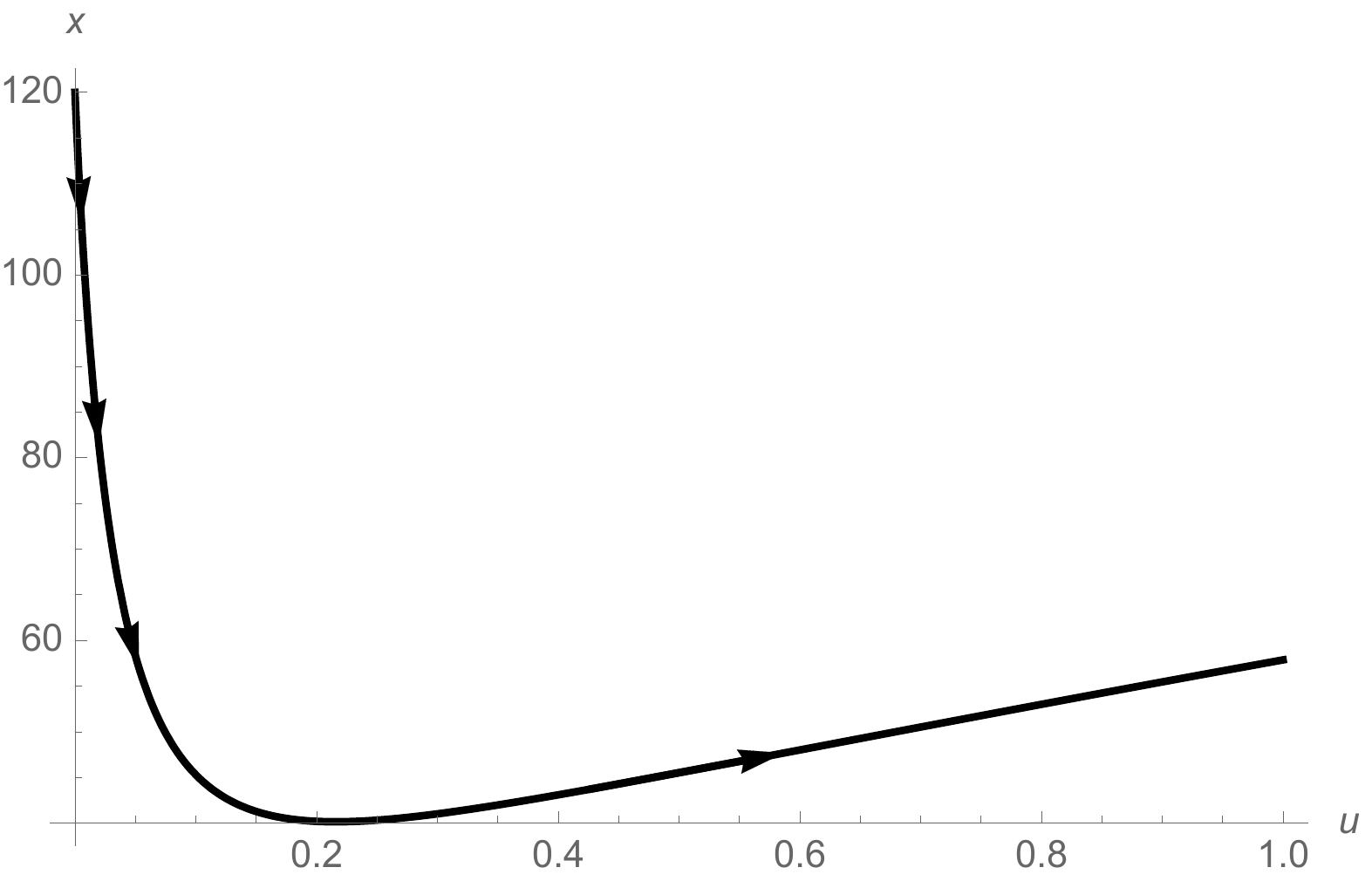}}
\caption{ $x$ from $120 \to 106.8451.$} 
\label{fig:runningx}
\end{subfigure}
\begin{subfigure}[htp]{.47\textwidth}
\centering
\setlength\fboxsep{4pt}
\fbox{\includegraphics[width=\textwidth]{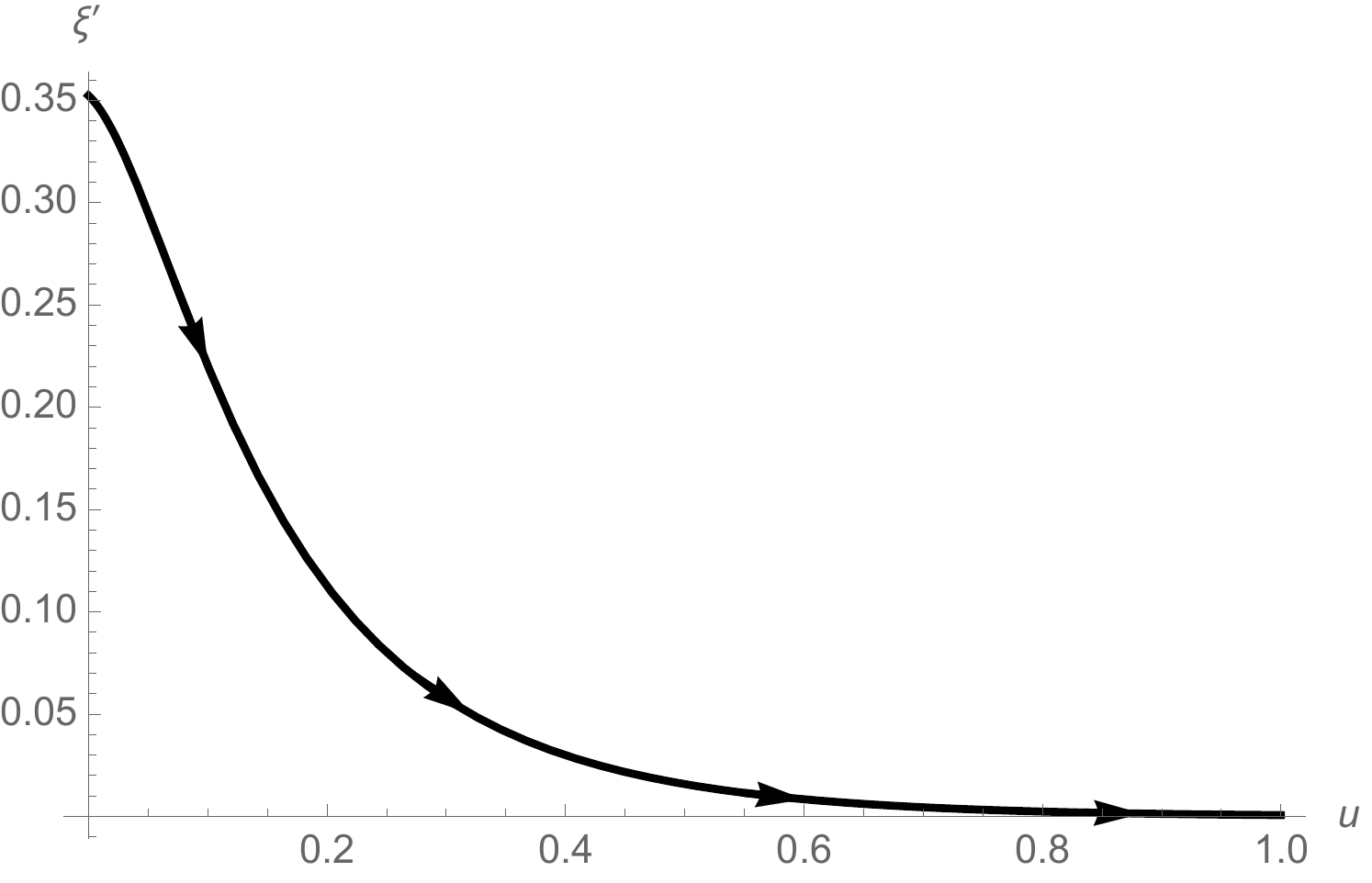}}
\caption{$\xi^\prime$ from $0.352307 \to -1.43995 {\times}10^{-5}.$} 
\label{fig:runningxip}
\end{subfigure}\\
\begin{subfigure}[htp]{.47\textwidth}
\hskip-.2in
\centering
\setlength\fboxsep{4pt}
\fbox{\includegraphics[width=\textwidth]{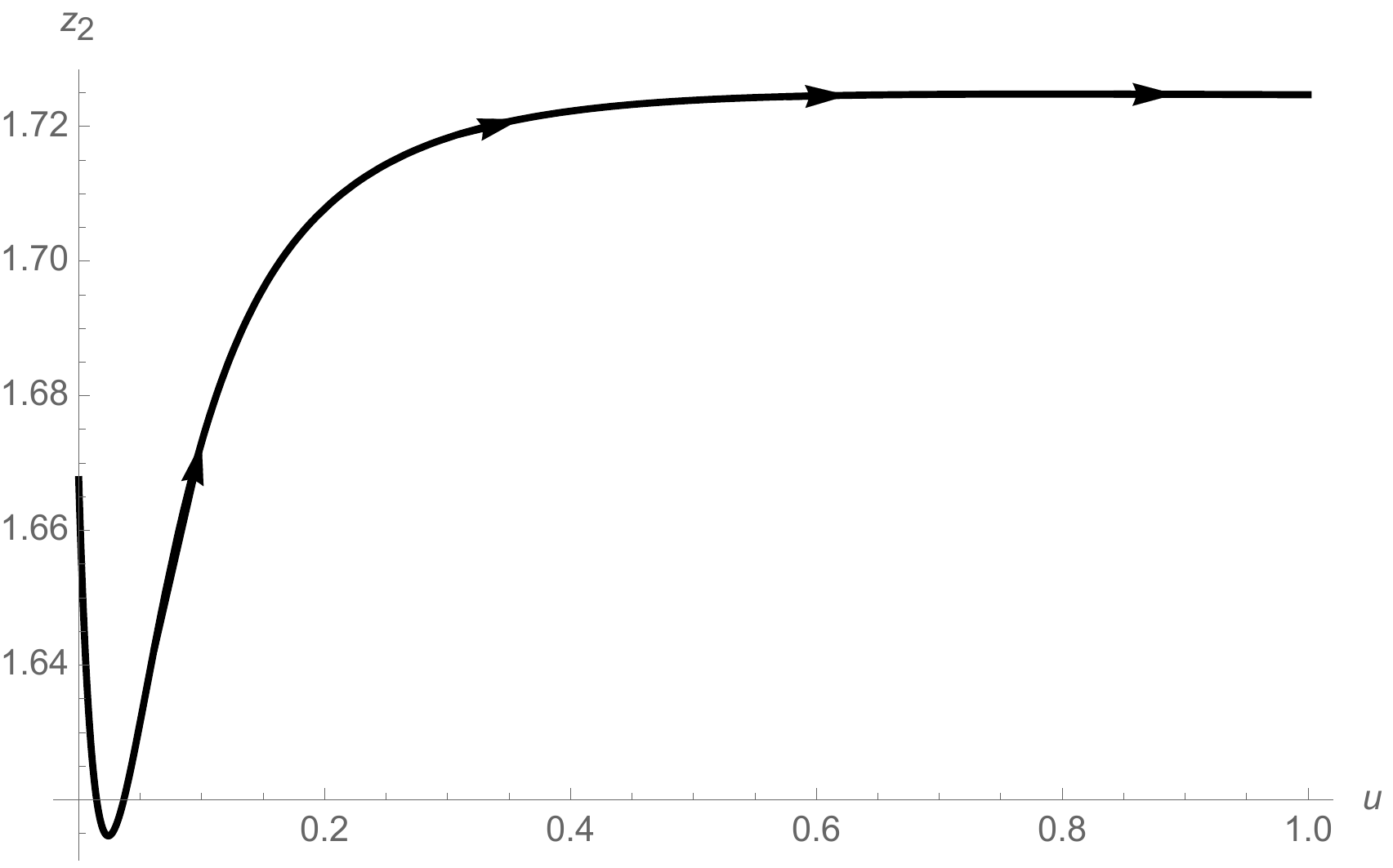}}
\caption{$z_2$ from $1.66754 \to 1.72354.$} 
\label{fig:runningz2}
\end{subfigure}
\begin{subfigure}[htp]{.47\textwidth}
\hskip-.12in
\centering
\setlength\fboxsep{4pt}
\fbox{\includegraphics[width=\textwidth]{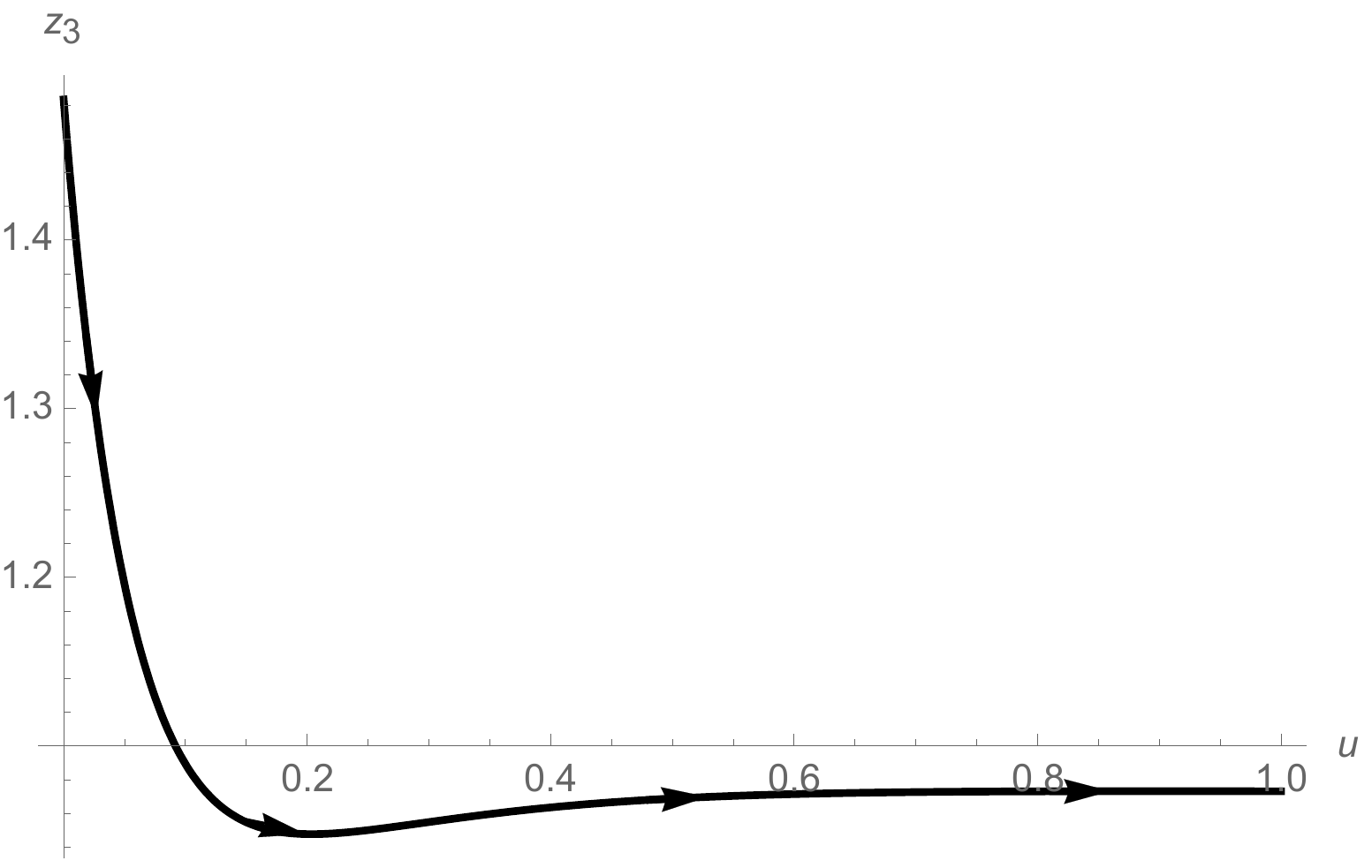}}
\caption{$z_3$ from $1.48330 \to 1.07062.$} 
\label{fig:runningz3}
\end{subfigure}
\caption{Running couplings up from a point on the DT-surface.}
\label{fig:running}
\end{figure}
All the other ratios $\{x, \xip, z_2, z_3\}$ mix with each other and 
with $\abar$, and it is far more difficult to determine their running 
analytically. Despite the complexity of these beta-functions in five 
variables, it is not difficult to solve for the running couplings numerically. 
In \figr{fig:running}, we present running couplings for one such 
case having the same parameters as in \figr{fig:dtsurface1} and 
\figr{fig:dtsurface2}. It is worth keeping in mind several of the basic 
parameters of this example: $T_F{=}41/2, N_a{=}330, b_g{=}2/3, 
b_2{=}791/20,$ not so very different from the example used in 
\secn{sec:betas}. From \figr{fig:dtsurface2}, we then selected a point at 
scale $v$ from the cross-hatched region from which to run: $(\abar {=} 
0.025, x {=} 120, \xip {=} 0.35231, z_2{=}1.66754,$ 
$z_3 {=}1.48330).$ 
These initial values and the associated UVFP from \tabby{tab:FP1} 
are given below each sub-figure. We display $\abar(u)$ 
running over a very large range of scales, but, to keep the figures of 
manageable size and to display the behavior near the DT-surface, 
only a small portion of the running is shown for the other four ratios.

We shall comment on some of the properties of these figures and use 
them as points of departure to summarize some of the qualitative 
features in other cases. The nonlinearity of the beta-functions is 
evident in many ways. For example, in \figr{fig:runningz2}, although the 
initial value of $z_2$ is not very far from its asymptotic value 
$z_2{}\uv$, it decreases rapidly at first before turning around and 
climbing back up. In other cases, where it starts at larger or smaller 
values, it may approach its UVFP far more directly. In \figr{fig:runningz3}, 
although $z_3$ starts above its UVFP, asymptotically it approaches it 
from below. In other cases within this same slice 
$(\abar{=}0.025, x {=} 120)$, it approaches from above. In 
\figr{fig:runningxip}, $\xip$ falls monotonically to its UVFP near zero, but 
one easily finds other cases where it rises initially before turning 
down. Finally, in \figr{fig:runningx}, we see that, although $x$ starts at 
$120$, not much larger than its UVFP value $\approx 107$, it falls 
dramatically to $\approx 40$ before turning upward again. That 
circuitous behavior is characteristic of this segment of the DT-surface, 
but it is not generic. When the initial value of $x\ll x\uv,$ we may find it 
increasing monotonically to its asymptotic value. Its behavior also is 
sensitive to whether one is in this ``stronger gravity" region, where 
initially $\abar{>}\abar{}\uv,$ or in the ``weaker gravity" region, where 
initially $\abar{<}\abar{}\uv.$ 

From the scale of \figr{fig:runningx}, it is not evident that $x$ ever grows to 
$x\uv.$ This is partly because we wanted to display the structure near the 
DT-surface but mostly because it runs much more slowly than the other 
couplings. The latter point is worth explaining. Note from 
\eqns{eq:betaxbar}{eq:betaxbar10} that $\betabar_x$ has a factor of $\abar$ in 
front. Because $\abar$ is small, $\betabar_x$ is relatively small, so that $x$ 
runs slowly. This behavior is the result of the conventional definition of $x$; this 
was one of the motivations for our introduction of $\bbar$ in \secn{sec:betas}. 
For the example presented in \tabby{tab:FP1}, the value of 
$\bbar{}\uv {=} x\uv \abar{}\uv {=} 1.80101.$ The corresponding figure that would 
replace \figr{fig:runningx} would show $\bbar$ running from $\bbar {=} 3.0$ on 
the DT-surface to $\bbar {=} 1.8,$ finally converging near its UVFP at the same 
rate as $\abar\to\abar\uv.$ 

Near the UVFP in our example, $\abar\uv/\bbar\uv \sim 10^{-2}.$ In 
\secn{sec:betas}, we showed that a small value is completely generic, 
this ratio depending only on $b_2,$ \eqn{eq:xuvaprx}. We may also check our 
estimate of $\xip\uv.$ For this model, the first term in 
\eqn{eq:betaxipbarp} takes the value $\approx -6.54,$ and 
$\wt{b}_g{=} 0.751,$ so that \eqn{eq:xipuv} yields 
$\overline{\Delta\beta}_\xip|_{\xip{=}0}\approx -5.79{<}0,$ negative, 
as required. Then from \eqn{eq:xipuv}, we get 
$\xip\uv\approx-1.51{\times}10^{-5},$ to be compared with the more 
precise value in \tabby{tab:FP1} of $-1.44{\times}10^{-5},$ only about 
a 5\% error.

Finally, we come to the UVFPs for $\{z_2,z_3\},$ approximated by the 
solutions of \eqn{eq:betazbar'}, which, for $SO(10),$ become
\begin{subequations}
\label{eq:betabbar10'/48}
\begin{align}
\label{eq:betaz2bbar10'/48}
\frac{\betabar{}_{z_2}}{48}& = \frac{3}{2}+
\frac{83}{5760}z_2^2+ 
\frac{1}{12} z_2 z_3 +\frac{\wt{b}_g z_2}{48} - z_2,\\
\label{eq:betaz3bbar10'/48}
\frac{\betabar{}_{z_3}}{48}& = \frac{3}{5} +
\frac{53}{144}z_3^2+ 
\frac{1}{2400}z_2^2 + 
\frac{1}{60}z_2 z_3+\frac{\wt{b}_g z_3}{48} - z_3.
\end{align}
\end{subequations}
Here we divided \eqns{eq:betaz2bbar'}{eq:betaz3bbar'} by the 
factor of $6(N{-}2)$ so as to normalize the coefficient of the negative 
contribution to one. With this normalization, we see that, the leading 
constants are $O(1),$ and the coefficients of the quadratic terms are 
all less than one. 

Let us first see how well these equations approximate the more 
precise solution in \tabby{tab:FP1}. In that case, we have 
$T_F \to 41/2, N_F \to 330,$ corresponding to 
$b_g \to 2/3, b_2 \to 791/20.$ Therefore, 
$\wt{b}_g \equiv b_g[1+5/b_2] {=} 594/791 \approx 0.751,$ 
significantly larger than $b_g.$ Solving simultaneously 
\eqns{eq:betaz2bbar10'/48}{eq:betaz3bbar10'/48}, we find that there are 
two FPs, of which one is a UVFP having the values 
$(z_2{=}1.7235, z_3{=}1.0706),$ agreeing to five significant figures with 
the values in \tabby{tab:FP1} calculated from the exact beta-functions! 
There is little doubt that this approximation captures the bulk of the 
effects due to dynamical gravity.

\begin{figure}[htbp]
\centering
\setlength\fboxsep{4pt}
\fbox{\includegraphics[width=4in]{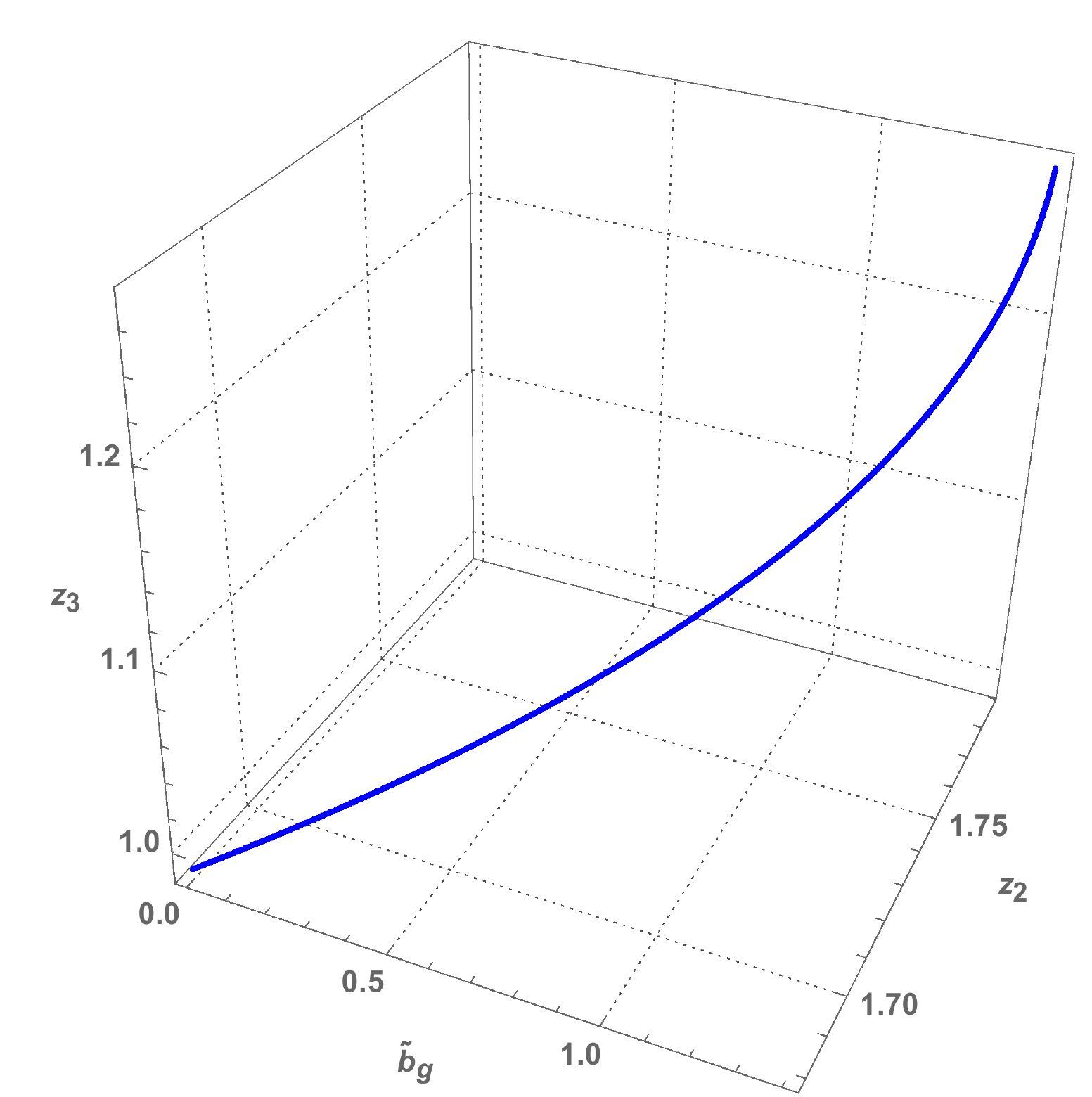}}
\caption{Potential range of UVFPs for $SO(10)$ with adjoint scalar.}
\label{fig:rangeuvfp}
\end{figure}

More generally, the simultaneous solution of 
$(\betabar{}_{z_2}{=}0, \betabar{}_{z_3}{=}0)$, 
\eqns{eq:betaz2bbar10'/48}{eq:betaz3bbar10'/48}, 
can be regarded as two constraints on the three parameters, 
$\{\wt{b}_g,z_2,z_3\}.$ Consequently, we can use this approximation 
to explore the range of solutions for all possible values of $\wt{b}_g.$ 
However, for reasons of stability and AF, we are only interested in 
solutions for which each parameter is positive. Solving numerically, we 
find UVFPs having positive values for the three parameters for the curve 
displayed in \figr{fig:rangeuvfp}. In particular, there are real positive solutions for 
$\{z_2,z_3\}$ only for $0{<}\wt{b}_g{<}1.406,$ and they range over 
$1.679{<}z_2{<}1.788$, $0.9884{<}z_3{<}1.273$. 

In a certain sense, the curve in \figr{fig:rangeuvfp} represents the entire range of 
conceivable UVFPs for the $SO(10)$ model with a single real adjoint scalar 
field. In reality, this simple model is much more restrictive. In \figr{fig:rangeuvfp}, we 
treated $\wt{b}_g$ as a continuous parameter, but of course, it only takes 
discrete values for the allowed values of $T_F,N_F.$ For $T_F=41/2,$ the 
range of $N_F$ depends on the choice of fermion representations. As we 
discussed below \eqn{eq:betabar10}, restricting fermions to the 
$\{ {\bf 10}, {\bf 16}, {\bf 45} \},$ there are 66 possible choices for $N_F,$ 
with $235 \leq N_F \leq 410,$ corresponding to 
$\sfrac{343}{10} \leq b_2 \leq \sfrac{861}{20},$ 
which in turn implies that $0.743{<}\wt{b}_g{<}0.762.$ 

As a second example, consider the case when $T_F{=}20$. In this case, 
$b_g{=}4/3$ and, one quickly determines that there is indeed a UVFP in the 
absence of dynamical gravity. To account for gravitational corrections, we need 
to replace $b_g$ with $\wt{b}_g$. With the restriction to the same fermion 
representations as before, there are again 66 cases with 
$225 {\leq} N_F {\leq} 400,$ corresponding to 
$\sfrac{343}{10} \leq b_2 \leq \sfrac{861}{20},$ and $1.488{<}\wt{b}_g{<}1.528.$ 
Even the minimum allowed value exceeds the upper limit of $\wt{b}_g{=}1.406.$ 
The effect of gravitational corrections has been to eliminate the 
UVFP\footnote{We have confirmed this conclusion with a more precise 
calculation using the exact beta-functions.}! Smaller values of $T_F$ (larger 
$b_g$) are obviously even worse. $T_F{=}41/2$ gives the only possible value of 
$b_g$ for which there is a UVFP for the scalar couplings! 

These examples illustrate the power of these approximations, enabling the 
determination of whether a UVFP exists for a model and, if so, providing rather 
accurate values for $\{\bbar\uv,\xi\uv,z_2\uv,z_3\uv\}$, together with calculable 
estimates of their uncertainties. 

As we have seen, the only place where 
nonlinearities become very important for estimating the UVFP is in 
\eqn{eq:betabbar10'/48}, which turns out to be extremely restrictive. We wish to 
conclude with a brief discussion of why that is. Because we must insist on 
finding solutions having positive $(z_2, z_3)$, these beta-functions have the feature 
that every term is positive except the linear term, ${-}z_k,$ which must offset 
the sum of all the other terms. As a result, the range of solutions is quite 
limited. The $z_k$ cannot be too small, because each formula, 
\eqns{eq:betaz2bbar10'/48}{eq:betaz3bbar10'/48}, has a constant term of 
$\ocal(1).$ If we completely ignore all the positive terms except for the 
constants, we quickly arrive at lower bounds of 
$\ocal(1)$: $z_2\uv{>}1.52, z_3\uv{>}0.61.$ At the same time, the solutions for 
$(z_2\uv, z_3\uv )$ cannot be too large because the quadratic terms will 
overwhelm 
the sole negative term in each beta-function. Just as $z\gtrsim z^2{>}0$ allows 
one to conclude $z\lesssim 1,$ one can make estimates of the upper limits 
coming from the quadratic terms here. Finally, $\wt{b}_g$ also contributes a 
positive, linear term that makes it even more difficult to have solutions. The 
upper limit on $\wt{b}_g$ may be far less than one might have guessed, but 
the ${+}\wt{b}_g z_k$ terms exacerbate a situation in which, even without them, 
it is already difficult to have AF scalar couplings.

\section{Conclusions and Outlook}
\label{sec:conclude}

We have succeeded in demonstrating within the context of a non-Abelian 
gauge theory coupled to renormalizable gravity that there exist 
regions of parameter space within which the three requirements listed 
at the conclusion of the introduction have been met: (1)~having 
AF with values of the coupling constants that ensure convergence 
of the EPI, (2)~manifesting DT perturbatively with a locally stable 
minimum, and (3)~lying within the catchment basin of the UVFP. We regard 
these three requirements as necessary for a sensible theory of this type.

Providing a renormalizable and AF completion of Einstein gravity, this 
model provides a connection between the Planck 
mass $M_P$, the cosmological constant $\Lambda$, the unification scale, 
$M_U \equiv \sqrt{\vev{T_2(\Phi)}} {=} r_0 \sqrt{\vev{R}/\alpha}
 {=} r_0 v/\sqrt{\alpha}$, 
the masses $M_V\sim r_0 v$ of the vector bosons, and the masses of heavy 
scalars arising from SSB. These relations are technically natural; the ratios 
of masses are functions of the coupling constants at scale $v$. It remains to 
explore in more realistic models how great a range of values result. 

To demonstrate local stability, 
we calculated the $O(\hbar^2)$ quantity $\varpi_2 \propto m_d^2,$ where 
$m_d$ is the mass of the dilaton~\cite{Einhorn:tbp}. We showed that 
there are regions of parameters space where $\varpi_2{>}0$ at the DT-scale, 
so extrema can be local minima. Since $m_d^2 {>} 0$ for some range of 
couplings, it may be that the usual conformal instability, characteristic of models 
starting from the Einstein-Hilbert action, is absent. This warrants further study.

Our discussion below \figr{fig:running} and elsewhere may make it sound 
as if, although the $SO(10)$ model is technically natural, a good deal 
of cooking has gone into the stew to make everything work out. In fact, we 
regard the need to follow a recipe as a positive aspect of this approach. 
The dynamical requirements dictate much about the choices of compatible 
representations of scalars and fermions. Indeed, the need for scalar 
couplings to be AF favors representations with large values of $T_F$ 
for fermions and even larger values of $T_S$ for scalars, so long as AF of 
the gauge coupling is maintained. Values of $b_g\lesssim \ocal(1)$ seem to 
be strongly favored. 

Given our limited goals for this paper, we have not included any mechanism 
describing further breaking of this symmetry down to the Standard Model. We 
hope it will be possible to do so, but it may be difficult to arrange that the 
splitting between the unification scale and the electroweak scale be naturally 
large. Perhaps there are supersymmetric extensions that would be technically 
natural, but we have not explored this possibility yet.

In principle, having overcome other limitations, we should now be in a
position to begin to investigate whether such models respect
unitarity. In the low energy theory below the DT scale, one might
be concerned with the possibility of a negative norm state, generally 
believed to be a problem for ``$R+R^2$'' gravity. The identification of
this issue relies on an expansion about flat space in order to write
the (inverse) quadratic form of the graviton fluctuations as
\beq
\frac{1}{M^2}\left(\frac{1}{k^2} - \frac{1}{k^2+M^2}\right)
\eeq
However, an inevitable consequence of DT is the existence of a
cosmological constant, so that flat space is not a solution to the
equations of motion. Thus the question is far more complicated than  it
might naively appear, dealing as it does with  spacetimes that are not
asymptotically flat, such as de~Sitter space. In fact, it has been known
for more than 30  years~\cite{Avramidi:1985ki, Avramidi:2000pia} that, 
to one-loop order, there are no unstable modes in de~Sitter background
provided the parameters of the model obey certain inequalities, which
our present model satisfies\footnote{There are however several zero
modes to be dealt with. This is reviewed in  \reference{Einhorn:tbp}}.
To our knowledge, this has been most thoroughly  investigated to by date
by Ashtekar, Bonga, and  Kesevan~\cite{Ashtekar:2015ooa,
Ashtekar:2014zfa, Ashtekar:2015lla, Ashtekar:2015lxa} who emphasize
several distinct features of de~Sitter space. No matter how small the
cosmological constant, there are ``no asymptotic Hilbert spaces in
dynamical situations of semi-classical gravity"~\cite{Ashtekar:2014zfa}.
Further, they show on physical grounds that one must include
non-normalizable growing modes among the gravitational waves on
$\cal{I}^+$. With all Killing fields spacelike at and near $\cal{I}^+$,
there is no way to define a conserved Hamiltonian and ``...~in the
quantum theory, we cannot decompose fields into positive and negative
frequency parts, even at $\cal{I}$~\ldots"~\cite{Ashtekar:2015lla}. It
seems as if the infrared problems in such spacetimes are more serious
than generally  believed and not simple generalizations from QED.
Theorems such as the  Ostrogradsky instability~\cite{Ostrogradsky:1850},
associated with Lagrangians containing higher than first-order time
derivatives\footnote{For recent discussions, see, e.g.,
\reference{Woodard:2015zca,Salvio:2015gsi, Langlois:2015skt}.}, would
seem not to apply.

Such spacetimes have no S-matrix and the attempts to generalize the ADM 
formalism to de~Sitter space are inadequate. Another property of spacetimes 
that are not asymptotically flat is that the G-B operator cannot be discarded. 
Although its coupling constant is determined by the other 
couplings up to a constant~\cite{Einhorn:2014bka}, it certainly played an 
important role in our derivation of the conditions for DT. It seems as if a new 
approach to QFTs with a positive cosmological constant may 
be required, both to resolve these theoretical challenges and to 
understand the observed ``Dark Energy".

Finally, the nature of measurement in theories with diffeomorphism 
invariance is complicated. It is conventional to say that there are no 
local, gauge-invariant observables. We take the view that, normally, 
this can be resolved once the measurement apparatus is included. 
Although the physical interpretation of a ``particle" is
frame-dependent, each piece of the experimental apparatus singles out a
special reference frame\footnote{This is no more anthropic than the
point of view of Gell-Mann and Hartle concerning decoherence, with
which we agree. See \reference{Gell-Mann:2013hza} and earlier papers
cited therein.}. Exactly how this is to be generalized to de~Sitter
space with strong curvature has not been precisely formulated. Although
correlation functions can be calculated in perturbation theory for any
particular choice of coordinates, without a Hilbert space and
well-defined norm, we are not quite sure how to define probability.
Until that has been spelled out, unitarity will probably remain an open
question.

Note also that in the high energy phase (where the
Higgs VEV is zero and there is no cosmological constant) the graviton
propagator has the form $1/k^4$ and it is an open question as to
whether this theory is physically sensible. We will discuss all these
issues at more length in a future publication,~\reference{Einhorn:tbp}.

The cosmological implications of these models, in particular, the
details of inflation, also remain to be developed but should be very
interesting. In a previous paper,~\cite{Einhorn:2012ih}, we
showed that the Higgs inflation 
paradigm~\cite{Bezrukov:2007ep,Barvinsky:2008ia}\ is in
fact compatible with a simple $SU(5)$ GUT structure, with the adjoint
Higgs being responsible both for inflation and the breaking to
$SU(3)\otimes SU(2) \otimes U(1)$. Aside from the issue of reassessing
this for the $SO(10)$ case, more difficult is the large value of $\xi$
associated with Higgs inflation. This large value caused controversy
regarding unitarity, but in our framework is clearly incompatible with
our use of perturbation theory at the DT scale, because of its effect on
the various dimensionless coupling $\beta$-functions, when gravity is
quantised.

Exactly what the nature of the medium is at scales much larger than $v$ 
is not at all obvious. Is it a plasma of particles or something else? 
Would it be possible to associate a temperature in this region? Is it hot or cold?

Without fine-tuning, $M_P, \Lambda, M_U,$ together with the dilaton mass 
$m_d,$ are all associated with a single scale $v,$ the scale of dimensional 
transmutation. This truly is a unification of gravity with particle physics. 
It appears as if the Big Bang may begin at the scale $v,$ which may be too large 
to explain the order of magnitude of inhomogeneities in the CMB. However, this
is only the beginning of an investigation into models of this type. It promises 
to be a very interesting development.


\acknowledgments
One of us (MBE) would like to thank A.~Vainshtein for discussions. 
DRTJ thanks KITP (Santa Barbara), the Aspen Center for Physics 
and CERN for hospitality and financial support.
This research was supported in part by the National Science Foundation under
Grant No. PHY11-25915 (KITP) and Grant No. PHY-1066293 (Aspen) and
by the Baggs bequest (Liverpool).

\begin{appendix}

\section{Lie algebra conventions}
\label{sec:liealgebra} 

In this paper, we limit ourselves to considering simple groups, mostly
$SO(N).$ The generalization to semi-simple groups is straightforward, 
since their algebras are the direct sum of the algebras of simple groups. 
$T_F, T_S$ are defined by the relation 
$\Tr[T^a T^b]\equiv T({\mathbf R})\delta^{a b}$ for any 
representation ${\mathbf R}$. In general, the representation 
${\mathbf R}$ will be reducible but expressible as the direct sum 
of irreducible representations.

For an irreducible representation, 
$\sum_a T^a T^a{=}C_2({\mathbf R}){\mathbf 1}_{d({\mathbf R})},$ 
where $d({\mathbf R})$ is the dimension of ${\mathbf R},$ and
$C_2({\mathbf R})$ is the quadratic Casimir invariant. It follows 
that $d({\mathbf R}) C_2({\mathbf R}) = d(G) T({\mathbf R})$. 
$\!C_G$ is equal to the quadratic Casimir $C_2({\mathbf G})$ for
the adjoint representation~${\mathbf G}.$

The precise relationship between $\{T({\mathbf G}), C_G\}$ and the 
gauge coupling $g$ depends upon the normalization convention 
for the generators. The convention in physics is $T_{\mathbf N}{=}1/2$ 
for the defining representation ${\mathbf N}.$ This choice gives for two classical series $C_{SO(N)}{=}(N{-}2)/2$, $C_{SU(N)}{=}N.$ 

For low-dimensional representations, 
this can be confusing. Even though the Lie algebras 
$SO(3) \cong SU(2),$ the fundamental for $SO(3), [SU(2)]$ is the 
vector ${\mathbf 3}\/$ [spinor ${\mathbf 2}$], respectively. 
For $SU(2),$ $C_{SU(2)}{=}C_2(\mathbf 3){=}2$; with 
$T({\mathbf 2})\equiv 1/2,$ then $C_2(\mathbf 2){=}3/4.$ 
For $SO(3),$ $C_{SO(3)}{=}C_2(\mathbf 3){=}1/2\equiv 
 T({\mathbf 3})$, so $T({\mathbf 2}){=}2C_2({\mathbf 2})/3{=}1/8$.

\section{Infrared Fixed Points}
\label{sec:IRFPs}

In this appendix, we explicate the analysis of the IRFPs of 
our $SO(10)$ model. 

Beginning again with $\abar,$ there are two possibilities for its IR
behavior, viz., depending on whether initially 
$\abar \to 0$ (weaker gravity region) or $\abar \to \infty$ (stronger gravity
region). In the first case, we may set $\abar {=} 0$ in these equations. 
Then $\betabar_x {=} 0$ at any fixed $x,$ and all dependence on $x$
drops out of the remaining beta-functions. In this case, $x$ is
undetermined. In fact, all gravitational corrections drop out in the
sense that all three $\Delta\beta_k {=} 0.$ Both $\betabar{}_{z_2}$
and $\betabar{}_{z_3}$ take their flat space values, and these
equations have two roots for $(z_2,z_3),$ one a UVFP and the other a
saddle for flat space. Inserting $\abar {=} 0$ and either of these values of
$(z_2, z_3)$ into $\overline{\beta_\xip}$, we see that
$\overline{\beta_\xip} {=} 0$ implies $\xip {=} 0,$ its conformal value. These
two FPs have been included in the text in 
\tabby{tab:FP1}, lines $5.^*$ \& $6.^*,$ because they
occur at finite $\abar.$ Since $\abar {=} 0$ is an IRFP for
$\betabar_\abar$, both solutions are in fact at best saddle 
in nature in the larger space; we called them ``saddle lines'' 
since $x$ is not determined at leading order.

As discussed above, the other possibility is $\abar \to \infty$ as
$t \to -\infty.$ This simply means that $a(t)$ increases faster than
$\alpha(t)$ in the IR and, in this case, our decision to rescale the
couplings by $\alpha$ does not serve us well. To determine the correct
behavior, we must re-express the beta-functions in terms of
$\alphabar \equiv \alpha/a {=} 1/\abar$ instead, and entertain the limit as
$\alphabar \to 0.$ At the same time, to seek IR fixed points, we must
introduce the rescaled parameter $du' \equiv a(t)dt$ and re-express
the beta-functions accordingly
\begin{subequations}
\label{eq:betapbar}
\begin{align}
\overline{\beta'}_\alphabar&=\alphabar\left( b_2-b_g \alphabar \right),\\
\overline{\beta'}&=\frac{1}{a}\beta=\alphabar \betabar\/\ {\mathrm{for}}\ 
\overline{\beta'}_x\ \/ {\mathrm{and}}\ \overline{\beta'}_\xip,\\
\left(\overline{\beta'}-b_2 z'\right)&=\alphabar^2\left( \betabar-b_g z \right)\ 
{\mathrm{for}}\ \overline{\beta'}_{z'_2}\/\ {\mathrm{and}}\ 
\overline{\beta'}_{z'_3},
\end{align}
\end{subequations}
where $z' {=} h/a {=} \alphabar z,$ for any of the scalar couplings.
We see that $\alphabar$ has a UVFP at $b_2/b_g,$ which is of course 
precisely the equivalent result as for $\abar,$ and the behavior of the couplings 
near there is just as before. As anticipated, however, it also has an IRFP at 
$\alphabar {=} 0.$

Using \eqn{eq:betapbar}, and with $(n_1,n_2,n_3) {=} (0,1,20)$ as before, 
$\overline{\beta'}$ may be expressed 
in terms of rescaled scalar couplings:
\begin{subequations}
\label{eq:betapbar10}
\begin{align}
\betabar'_x&=
\left[-\frac{10}{3}+\frac{891}{20} x-
\left(\frac{5}{12}+\frac{135}{2}\xip^2\right)x^2\right],
\label{eq:betax10'}\\
\betabar'_\xip&=\left(\frac{2}{5}z'_2+\frac{47}{3}z'_3-24\alphabar\right)\xip+
\left(\xip-1/6\right)\left(\frac{10}{3x}-\frac{3}{2}x \xip(2\xip+1)\right),\\
\betabar'_{z'_2}&=72\alphabar^2+\frac{83}{120}z'_2{}^2+ 
4z'_2z'_3+\left(\frac{791}{20} -\frac{142}{3}\alphabar\right) z'_2+
\left(5-18 x \xip^2\right)z'_2,\\
\begin{split}
\betabar'_{z'_3}&=\frac{144}{5}\alphabar^2+\frac{53}{3}z'_3{}^2+ 
\frac{1}{50}z'_2{}^2+\frac{4}{5}z'_2z'_3+\left(\frac{791}{20} - 
\frac{142}{3}\alphabar\right)z'_3+\\
&\hskip25mm 
\left(5-18 x \xip^2\right)z'_3+3(\xip-1/6)^2\left( 5+9x^2\xip^2 \right).
\end{split}
\end{align}
\end{subequations}
Note that $\betabar'_x$ is independent of $\alphabar$ and has two FPs 
in $x$ for fixed $\xip,$ the same two as it had previously for $\abar\neq 0$:
\beq
\label{eq:xpm}
x_\pm=\frac{2673\pm\sqrt{7124929-(1800 \xip)^2}}{50 (1+162 \xip^2)}
\eeq
The larger one $x_+,$ is a candidate UVFP in $x$; the smaller, $x_-,$ a 
candidate IRFP. Of course, this only makes sense if there is a FP for $\xip$ in 
the range $0 \leq |\xip| {<} \sqrt{7124929}/1800,$ which we shall find is a correct 
assumption. Taking $\alphabar \to 0$ in the remaining equations, the other three 
$\overline{\beta'}$ are
\begin{subequations}
\label{eq:betapbar100}
\begin{align}
\betabar'_\xip& {=} \left(\frac{2}{5}z'_2+\frac{47}{3}z'_3\right)\xip{+}
\left(\xip{-}1/6\right)\left(\frac{10}{3x}-\frac{3}{2}x \xip(2\xip+1)\right),\\
\betabar'_{z'_2}& {=} z'_2\left[\frac{83}{120}z'_2+4z'_3+\frac{891}{20} - 
18 x \xip^2\right],\\
\betabar'_{z'_3}& {=} \frac{53}{3}z'_3{}^2+\frac{1}{50}z'_2{}^2+ 
\frac{4}{5}z'_2z'_3+\!\left(\frac{891}{20} -18 x \xip^2\right)z'_3{+}
3(\xip{-}1/6)^2\left( 5+9x^2\xip^2 \right)\!.
\end{align}
\end{subequations}
These take the form of the theory without gauge interactions (even 
though both $a$ and $\alpha$ blow up in this limit.) 

\begin{table}[htbp]
\begin{center}
\begin{tabular}{|c|c|c|c|c| c| c| } \hline
& $ \alphabar $ & $ x $ & $ \xip $ & $ z'_2$ & $z'_3$ & Nature\\ \hline
$\!{\bf 1.} $ & $ {\bf 0} $ & {\bf 0.0751125} & ${\bf 0.166667} $ & $ {\bf 0} $
& $ {\bf 0} $ & {\bf IRFP} \\ \hline
$2.\ $ & $ 0 $ & $ 19.3649$ & $~0.166667$ & $ 0 $ & $ 0 $ & saddle point\\ 
\hline
$3.\ $ & $ 0 $ & 0.0931182 & $ 1.17747 $ & $ 0 $ & $ -1.93125 $ 
& saddle point \\ \hline
$4.\ $ & $ 0 $ & $ 69.5124 $ & $ -0.0575469 $ & $ 0 $ & $-1.33242$ 
& saddle point\\ \hline
$5.\ $ & $ 0 $ & $ 106.842 $ & $-4.13596{\times}10^{-4}$ & $ 0 $ & $ 
-2.51221 $ 
& saddle point \\ \hline
$6.\ $ & $ 0 $ & 106.844 & \ \ $ 1.95663{\times}10^{-4} $ & $ 0 $ & 
$ -0.00937304 $ 
& saddle point\\ \hline
\end{tabular}
\caption{Infrared fixed points for an $SO(10)$ model for 
$\alphabar {=} 0$ $(\abar \to \infty).$}
\label{tab:FP2}
\end{center}
\end{table}

We remarked earlier that, 
without the gauge coupling, the scalar couplings in \eqns{eq:betas} 
{eq:betabars} have no UVFP, and that is the case here as well. On the other 
hand, they may well have other FPs. Setting each of these equal to zero, we 
can solve. We list the results in \tabby{tab:FP2}. It is amusing that one of 
the FPs at $\alphabar {=} 0$ $(\abar \to \infty)$ is an IRFP, but of course, this 
perturbative calculation is not trustworthy in that limit.

\section{Stability of classical extrema}
\label{sec:var2}

In this appendix, we include details concerning the classical 
stability of the extrema of the action, \eqn{eq:action2}. To review, 
the extrema take the form 
$\vev{\varphi^{[k]}}=r^{[k]}\mathrm{Diag}(\omega_k,-\omega_k),$ $\{k{=}0,{\ldots},4\}$, as described in the discussion surrounding \eqns{eq:classicalssb}{eq:r0k}. Then 
$\vev{t_2}=r^{[k]}{}^2(10{-}2k),$ 
$\vev{\varphi^{[k]}{}^3}=r^{[k]}{}^2\vev{\varphi^{[k]}},$ and 
$\vev{\varphi^{[k]}{}^4}=r^{[k]}{}^2 \vev{\varphi^{[k]}{}^2}$. 
%
The second variation, or ``stability matrix," is given in \eqn{eq:var2}; as noted earlier, it involves four distinct traces. Going on-shell by replacing 
$\varphi \to \vev{\varphi^{[k]}}$, they take the values
\begin{subequations}
\begin{align}
\Tr[\delvar^2]&=2\textstyle{\sum_{mn}}\left[\left|(\delvar_1)_{mn}\right|^2{+}\left|(\delvar_2)_{mn}\right|^2\right],\\
\Tr[\varphi\delvar]&=2r^{[k]}\Tr[\omega_k\delvar_1]=
2r^{[k]}\textstyle{\sum^{\prime}_{mm}}(\delvar_1)_{mm},\\
\begin{split}
\Tr[\varphi^2\delvar^2]&=r^{[k]}{}^2\left[2\Tr[\omega_k\delvar_1^2]+\Tr[\omega_k\{\delvar_2,\delvar_2^\dag\}]\right]=\\
&\hspace{5mm} 2r^{[k]}{}^2\textstyle{\sum^{\prime}_{mn}}\left[\left|(\delvar_1)_{mn}\right|^2{+}
\left|(\delvar_2)_{mn}\right|^2\right],
\end{split}\\
\Tr[(\varphi\delvar)^2]&=2r^{[k]}{}^2\textstyle{\sum^{\prime}_{mn}}\left[\left|(\delvar_1)_{mn}\right|^2{-}
\left|(\delvar_2)_{mn}\right|^2\right],
\end{align}
\end{subequations}
where the prime on the summation in the last three formulas denotes restricting $m$ to the non-null components of $\omega_k$ (but summing over all $n$.) Pulling these pieces together, \eqn{eq:var2} becomes
\begin{align}\label{eq:delta2k}
\begin{split}
\frac{\delta^2S_{cl}^{(os)}}{V_4}&=r^{[k]}{}^2\left\{\frac{4h_1}{3}
\left|\textstyle{\sum^{\prime}_{mm}}(\delvar_1)_{mm}\right|^2
+\frac{h_2}{12}\textstyle{\sum^{\prime}_{mn}}\left[
3\left|(\delvar_1)_{mn}\right|^2{+}
\left|(\delvar_2)_{mn}\right|^2\right]\right\}\\
&+\left(\frac{h_1r^{[k]}{}^2(10{-}2k)}{3}-2\xi\right)
\textstyle{\sum_{mn}}\left[\left|(\delvar_1)_{mn}\right|^2{+}\left|(\delvar_2)_{mn}\right|^2\right].
\end{split}
\end{align}
Noting \eqn{eq:classicalssb}, we can write the coefficient of the last term as 
${-}h_2 r^{[k]}{}^2/12,$ in which form, it is simpler to combine with the other terms having coefficient $h_2.$ However, to do so requires breaking up the sum into the restricted sum $\sum^{'}$ plus the remaining terms $\sum^{''}.$
Then the second variation \eqn{eq:delta2k} becomes 
\begin{align}\label{eq:delta2kp}
\begin{split}
\frac{\delta^2S_{cl}^{(os)}}{V_4}&=\frac{r^{[k]}{}^2}{3}\left\{4h_1
\left|\textstyle{\sum^{\prime}_{mm}}(\delvar_1)_{mm}\right|^2
+h_2\textstyle{\sum^{\prime}_{mn}}\left[
\left|(\delvar_1)_{mn}\right|^2\right]\right\}-\\
&\hspace{5mm} \frac{h_2 r^{[k]}{}^2}{12}\textstyle{\sum^{''}_{mn}}
\left[\left|(\delvar_1)_{mn}\right|^2{+}\left|(\delvar_2)_{mn}\right|^2\right].
\end{split}
\end{align}
We see from the second term in the first line that the off-diagonal 
contributions to $\delvar_1$ are stable only if $h_2{>}0$. On the other hand, 
the second line (involving $\sum^{''}$) restricts $m$ to be in 
the null subspace of $\vev{\varphi}.$ As a result, this sum contains 
fluctuations $\{(\delvar_1)_{mn},(\delvar_2)_{mn}\}$ that occur nowhere else 
in \eqn{eq:delta2kp}, and, since they enter with a minus sign, such 
fluctuations are stable only for $h_2{<}0.$ Thus, for either sign of $h_2,$ 
there is an instability.

Consequently, the only possibility of finding a nontrivial, stable minimum is for 
the case $k{=}0,$ when $\vev{\varphi}$ has no zero eigenvalues and the 
second line is absent. In that case, the preceding equation simplifies to 
\begin{align}\label{eq:delta2zero}
\frac{\delta^2S_{cl}^{(os)}}{V_4}&=\frac{r^{[0]}{}^2}{6}\left\{8h_1
\left|\textstyle{\sum_{m}}(\delvar_1)_{mm}\right|^2
+h_2\textstyle{\sum_{mn}}\left[
\left|(\delvar_1)_{mn}\right|^2\right]\right\}.
\end{align}
$\delvar_2$ drops out, so those fluctuations do not get mass. These are the 
would-be Goldstone bosons that, in the gauge theory, get ``eaten" to form the 
massive vectors. The remaining fluctuations are the $SU(5){\otimes}U(1)$ 
invariant scalars that get masses. For stability, so that these particles are not tachyons, this expression must be nonnegative for all fluctuations $\delvar_{mn}$. The off-diagonal elements 
contribute $\propto h_2\sum_{n>m}|\delvar_1{}_{mn}|^2,$ so we must have 
$h_2>0$. 



Setting the off-diagonal elements zero, the diagonal elements of $\delvar_1$ 
make up a homogeneous polynomial of degree two in five real variables. For 
$\delvar_1$ diagonal , we can rewrite the curly brackets in \eqn{eq:delta2zero} as
\beq
\left(8h_1+h_2/5\right)\Tr[\delvar_1]^2+
h_2\Tr\!\left[\left(\!\delvar_1-\frac{\Tr[\delvar_1]}{5}\right)^{\!\!2}\right].
\eeq
Therefore, this is nonnegative provided $h_1{+}h_2/40=h_3>0.$ In sum, this symmetry-breaking is stable provided both $h_2, h_2$ are positive.

\section{DT Scale and Stability}
\label{sec:dtscale}

The formulas for the determination of the DT scale $v$ and the nature 
of the extrema are simple in principle but quite complicated in practice, 
even in the oversimplified model of matter considered in this paper. 
For completeness, 
we present the formulas for the on-shell values of 
$B_1^{(os)}$ and $\varpi_2^{(os)}$ for $SO(10)$ with an arbitrary 
number $N_F$ of fermions and with contribution $T_F$\/ to the gauge 
boson beta-function.


\begin{align}
\begin{split}
\label{eq:b1osfull}
 \hskip-7mm B_1^{(os)}&=\frac{11N_f{-}5026}{4320}{+}
\frac{10}{9 x^2}{-}\frac{5}{3 x}{+}\xip(2\xip{-}1){+}
\frac{\left(6\xip{-} 1\right)(60{-}z_2)}{30 z_3}+\hskip15mm \\
&\hskip5mm
\frac{\left(6\xip
{-} 1\right)^2(z_2^2{+}1440)}{1200 z_3^2} {+}
\frac{\abar(6\xip{-} 1)^2\left(20{-}15 x{+}
9 x^2 \xip \left(4 \xip{-}1\right)\right)}{72 x z_3}+\\
&\hskip5mm
\frac{\abar^2(6\xip{{-}} 1)^4(5{+}9 x^2\xip^2)}{288 z_3^2},
\end{split}
\end{align}
\nopagebreak[1]
\begin{align}
\begin{split}
\label{eq:varpi2osfull}
\varpi_2^{(os)}&{=}\frac{\alpha}{2 160 000 x^3 z_3^3}
\bigg[ {-} 625\, \abar^4 x^3 ( 6 \xip {-} 1)^6 \left(5 {+} 
9 x^2 \xip^2\right)^2 - \cr
&1875\, \abar^3 x^2 z_3 ( 6 \xip {-} 1)^4 \bigg(\! {-} 200 {+} 
x \Big(611 {+} N_F{+} 110 x \xip {-} 720\, x\xip^2+ \cr
&
3 x^2\xip^2 \big( {-} 2427 {+} 2 x (1 {+} 33 \xip {+} 
8820 \xip^2) \big)\Big)\!\bigg) 
+ \cr
&
150\, \abar^2 x (6 \xip{-} 1)^2 
\bigg[ {-} 2 x^2 (6 \xip{-} 1)^2 \left(5 {+} 9 x^2 \xip^2\right) \left(1440 {+} z_2^2\right) + \cr
&
60\,x^2 ( 6 \xip{-} 1 ) \left(5 {+} 6 x^2 \xip^2 (1 {+} 3 \xip)\right) 
( z_2{-} 60 ) z_3 - \cr
&
25 z_3^2 \bigg(\! 800 {+} x \Big( ( 3 x {-} 8 )N_F {+} 
1913 x{+} x \xip \Big(7191 x ( 4 \xip {-} 1 ) {+} \cr
&40 ( 2907 \xip{-} 22 ) {+}
6 x^2 \big(1 {+} 4 \xip (8 {+} 3 (769 {-} 3222 \xip) \xip) \big)\Big){-} 
21268 \Big)\! \bigg) \bigg] 
+ \cr
&
100\, \abar z_3 \bigg(\! 3 x^2 (6 \xip{-} 1)^2 
\Big(1440 \Big(40 {+} 9 x \big( {-} 5 {+} 
2 x \xip ( 7 \xip {-} 1 ) \big) \! \Big) + \cr
&
z_2^2 \big(3 x ( 6 x ( \xip {-} 1)\xip {-} 5 ){+} 40 \big)\! \Big)\! {-} 
360\, x^2 z_3 ( 6 \xip{-} 1 ) \Big(\! 10 (z_2{-} 60 ) + \cr
&
x \left(600 {-} 5 z_2 \right){+}3 x^2 \xip \big(60 {+} 24 \xip^2
( z_2 {-} 60 ) {-} z_2 {-} 4 \xip (30 {+} z_2) \big) \Big) {-} \cr
& 50 z_3^2 \Big( 3 x \Big(37648 {+} (8 {-} 6 x)N_F {+} 12 x^3 \xip
( 6 \xip{-} 1 ) ( 2 \xip ( 411 \xip{-} 85) {-} 3 ) + \cr
&
45 x^2 \big(7 {+} 4 \xip ( {-} 2 {+} 949 \xip)\big) {-} 
4 x \big(7099 {+} 20 \xip ( 2859 \xip{-} 11)\big)\!\Big) 
{-} 1600 \Big)\! \bigg) + \cr
&
3 x^3 \Big[ 5 z_3 ( 6 \xip{-} 1 ) 
\Big( 230400 ( (6\xip{-}1)( T_F{-}3) {-} 36 ) {-} 8640 z_2 ( 42 \xip {-} 23) {-} \cr
&
2880 (1 {+} 6 \xip) z_2^2 {+} (61 {+} 210 \xip) z_2^3 \Big) {+} 
100 z_3^2 \Big(1920\left( 6 (69 {+} 5 T_F) \xip {-} 
324 \xip^2{-} \right. \cr
&\left. 5 (30 {+} T_F) \right) + 5760 z_2 {+} ( 6 \xip (35 {+} 72 \xip){-} 95) z_2^2\Big){-} 
12 (6 \xip{-}1)^2 \left(1440 {+} z_2^2\right)^2 + \cr
&
4000 \big(3180 {+} 72 \xip (2 \xip ( z_2{-}60) {-} z_2) {-} 41 z_2\big) 
z_3^3 {+} 5\,640\, 000 \xip ( 4 \xip {-} 1 ) z_3^4 \Big] \bigg].
\end{split}
\end{align}

\end{appendix}

\vfill
\newpage


\end{document}